\documentclass[sigconf,10pt]{acmart}

\setcopyright{rightsretained}
\AtBeginDocument{%
  }

\usepackage{tikz}
\usepackage{amsmath}
\usepackage{bm}
\usepackage{mathptmx} % This is Times font
\usepackage{booktabs}

\usepackage{bbding}                                        
\usepackage{ulem}
\usepackage{endnotes,microtype,xspace,graphicx,fancyvrb,multirow}
\usepackage{array}
\usepackage{enumitem}
\usepackage{balance}
\usepackage{xcolor}
\definecolor{darkred}{rgb}{0.8, 0.1, 0.1}
\usepackage{xparse}

\usepackage{textcomp}
\usepackage{pifont}
\usepackage{fontenc}
\usepackage{bbding}  
\usepackage[vlined,ruled,linesnumbered]{algorithm2e}
\usepackage{graphicx}
\usepackage{pythonhighlight}
\usepackage{svg}

\newcommand{\sys}{\textsc{Jiagu}\xspace}
\newcommand{\ssim}{S${^3}$imulator\xspace}

\newcommand{\hwcloud}{Huawei Cloud\xspace}
\newcommand{\concurrency}{highly-replicated\xspace}
\renewcommand{\emph}[1]{\textit{#1}}

\newcommand{\myparagraph}[1]{\vspace {3pt}\noindent\textbf{\emph{#1}}}
\newcommand{\TODO}[1]{\textcolor{red}{TODO: #1}}

\newcommand{\DD}[1]{\textcolor{blue}{Dd: #1}}
\newcommand{\LQY}[1]{\textcolor{violet}{LQY: #1}}

\newcommand{\MZY}[1]{\textcolor{purple}{MZY: #1}}

\newcommand{\fa}{$\textit{f1}$}
\newcommand{\fb}{$\textit{f2}$}
\newcommand{\fc}{$\textit{f3}$}
\newcommand{\cached}{cached\xspace}
\newcommand{\saturated}{saturated\xspace}

\newcommand{\highlight}[1]{\textcolor{violet}{#1}}
\newcommand{\release}{release\xspace}

\definecolor{hightcode}{rgb}{1.0,0.13,0.32}
\newcommand{\blue}[1]{\textcolor{blue}{#1}}
\newcommand{\green}{\color[HTML]{008000}}
\newcommand{\red}[1]{\textcolor{red}{#1}}
\newcommand{\tabincell}[2]{\begin{tabular}{@{}#1@{}}#2\end{tabular}}
\newcommand{\cmark}{\ding{51}}
\newcommand{\xmark}{\ding{55}}

\NewDocumentCommand{\codeword}{v}{\emph{#1}}

\begin{document}

\title{\sys: Optimizing Serverless Computing Resource Utilization with Harmonized Efficiency and Practicability}

\author{Qingyuan Liu}
%\authornote{This is a note.}
\affiliation{%
  \institution{Shanghai Jiao Tong University}
  \state{Shanghai}
  \postcode{200240}
  \country{China}
}

\author{Yanning Yang}
%\authornote{This is a note.}
\affiliation{%
  \institution{Shanghai Jiao Tong University}
    \state{Shanghai}
  \postcode{200240}
  \country{China}
}
% \email{liu\_qy@sjtu.edu.cn}

\author{Dong Du}
\affiliation{%
  \institution{Shanghai Jiao Tong University}
  \state{Shanghai}
  \postcode{200240}
  \country{China}
}
% \email{dd_nirvana@sjtu.edu.cn}

\author{Yubin Xia}
\affiliation{%
  \institution{Shanghai Jiao Tong University}
  \state{Shanghai}
  \postcode{200240}
  \country{China}
}
% \email{xiayubin@sjtu.edu.cn}

\author{Ping Zhang}
\affiliation{%
  \institution{Huawei Cloud}
  \country{China}
}

\author{Jia Feng}
\affiliation{%
  \institution{Huawei Cloud}
  \country{China}
}

\author{James Larus}
\affiliation{%
  \institution{EPFL}
  \country{Switzerland}
}
% \email{pingzhang8-c@my.cityu.edu.hk}

\author{Haibo Chen}
\affiliation{%
  \institution{Shanghai Jiao Tong University}
  \state{Shanghai}
  \postcode{200240}
  \country{China}
}

% \setlength{\affilsep}{0.5em}

% \author{Qingyuan Liu$^{1,2}$, Dong Du$^{1,2}$, Yubin Xia$^{1,2,3}$, Ping Zhang$^4$, Feng Jia$^4$, James Larus $^5$, Haibo Chen$^{1,2}$}

% \affiliation{
%   \institution{
%   {$^1$Institute of Parallel and Distributed Systems, Shanghai Jiao Tong University} \\
%   {$^2$Engineering Research Center for Domain-specific Operating Systems (MoE)} \\
%   {$^3$Shanghai AI Laboratory} \\
%   {$^4$Huawei Cloud} \\
%   {$^5$EPFL}
%   }
%   \country{}
% }

\newcommand{\cleanauthors}{Qingyuan Liu, Dong Du, Yubin Xia, Ping Zhang, Haibo Chen}
% \author[1]{Qingyuan Liu}
% \author[1]{Dong Du}
% \author[1,2]{Yubin Xia}
% \author[3]{Ping Zhang}
% \author[1]{Haibo Chen}
% \affil[1]{Institute of Parallel and Distributed Systems, Shanghai Jiao Tong University, Engineering Research Center for Domain-specific Operating Systems (MoE)}
% \affil[2]{Shanghai AI Laboratory}
% \affil[3]{Huawei Cloud}

% \email{haibochen@sjtu.edu.cn}

%%
% \settopmatter{printfolios=true}
% \settopmatter{printacmref=false} 
% \setcopyright{none}
% \renewcommand\footnotetextcopyrightpermission[1]{}
\begin{abstract}
% \highlight{The difficulty in harmonizing the effectiveness and efficiency has hindered serverless computing from proficiently utilizing resources, and therefore we propose \sys.}
% Overcommitment requires predicting performance to prevent unacceptable performance degradation and QoS violation, introducing trade-off between prediction accuracy and overheads.

Current serverless platforms struggle to optimize resource utilization due to their dynamic and fine-grained nature.
Conventional techniques like overcommitment and autoscaling fall short, often sacrificing utilization for practicability or incurring performance trade-offs.
Overcommitment requires predicting performance to prevent QoS violation, introducing trade-off between prediction accuracy and overheads.
Autoscaling requires scaling instances in response to load fluctuations quickly to reduce resource wastage, but more frequent scaling also leads to more cold start overheads.
This paper introduces \sys to harmonize efficiency with practicability through two novel techniques.
First, \emph{pre-decision scheduling} achieves accurate prediction while eliminating overheads by decoupling prediction and scheduling.
Second, \emph{dual-staged scaling} achieves frequent adjustment of instances with minimum overhead.
We have implemented a prototype and evaluated it using real-world applications and traces from the public cloud platform.
Our evaluation shows a 54.8\% improvement in deployment density over commercial clouds (with Kubernetes) while maintaining QoS, and 81.0\%--93.7\% lower scheduling costs and a 57.4\%--69.3\% reduction in cold start latency compared to existing QoS-aware schedulers in research work.

\end{abstract}

\maketitle

\pagestyle{plain}

\section{Introduction}

\iffalse %% Commented!
In serverless computing~\cite{jonas2019cloud}, the responsibility of Dev/ops is shifted from users to providers~\cite{AWSLambda, IBMcloudFn, MSFunc, GoogleFunc}.
The provider deploys mulitple homogeneous instances to execute a single purpose user function.
Each instance is allocated a user-specified amount of resources.
User requests of a function is firstly sent to a router, which commonly applies load balance policy and tries to evenly send the requests to the function's instances for processing. 
Before creating an instance, a scheduler would choose an appropriate node to deploy it.
Another feature of serverless computing is autoscaling. 
A specific value of loads (i.e., RPS, requests per second) is defined for each function as its ``saturated value'', and the autoscaler creates new instances of a function when the average load of the function's instances reaches the saturated value~\cite{knative-scalar,openfaas-scalar}, helping to avoid instance overloads\footnote{It means that loads per instance would commonly not exceed the saturated value, as exceeding that value would result in creating new instances, which splits some of the load.} and guarantee performance.
Commonly, the performance requirements are quantified by the Quality-of-Service (QoS).
%\TODO{Refine the above background.}
\fi

Serverless computing~\cite{jonas2019cloud} simplifies the management of cloud applications with features like on-demand execution and autoscaling.
Through autoscaling, cloud providers are able to dynamically deploy zero to numerous instances to run a specific function in response to real-time demands, optimizing performance and cost efficiency.
Critical to this process are two main components: the autoscaler and the scheduler.
The autoscaler utilizes a predefined threshold for each function, called \emph{saturated value} (e.g., requests per second, or RPS), and triggers the creation of new instances as needed when the live load approaches this threshold~\cite{knative-scalar,openfaas-scalar}\footnote{This approach typically ensures individual instance loads do not exceed the saturated value, as doing so would lead to the instantiation of additional instances, thereby redistributing the load.}.
By doing so, the autoscaler prevents instances from becoming overloaded and guarantees performance, which is generally measured in terms of Quality-of-Service (QoS).
When the autoscaler triggers the creation of a new instance, the scheduler is responsible for scheduling the instance to a suitable node.
In making these assignments, the scheduler accounts for various aspects, including whether a node can meet a function's resource requirements which developers manually define.
% As a result, serverless has been widely adopted by nowaday cloud vendors~\cite{AWSLambda, IBMcloudFn, MSFunc, GoogleFunc}.

% The number of instances are deterimned by the loads.

\begin{figure}[t]
  \centering
 \setlength{\belowcaptionskip}{-10pt}
  \setlength{\abovecaptionskip}{-1pt}
   \begin{minipage}[t]{0.98\linewidth}
    \centering
    \includegraphics[width=0.78\textwidth]{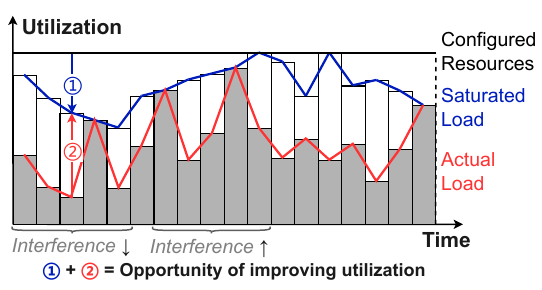}
    \footnotesize
    \end{minipage}
    \caption{\textbf{Opportunities of improving resource utilization.} 
    ``Configured resources'' means the allocated resources can be fully utilized.
    The blue line demonstrates the resource utilization of instances that are serving a saturated value of load, which fluctuates due to the variation in resource interference.}
%\MZY{The x and y axis should use scheduling costs and Instance QoS.}\DD{fixed.}
  \label{fig:util-motiv}
\end{figure}

However, current serverless systems still fall short on efficiently utilizing resources.
According to traces from AliCloud~\cite{10.1145/3542929.3563470} and our public cloud (i.e., \hwcloud, a top-five IaaS provider~\cite{gartner-2021-market}), most serverless functions can only utilize a small portion of allocated resources.
%(Figure~\ref{fig:motiv-util-cdf}).
As illustrated in Figure~\ref{fig:util-motiv}, the issue mainly results from two factors.
First, to guarantee performance, users usually consider the worst case, and thus specify excessive resources to instances.
Even if instances are processing saturated load of requests, they cannot utilize the allocated resources.
This causes the resource wastage of part \blue{\ding{172}} in Figure~\ref{fig:util-motiv}.
Second, in practice, user loads served by each instance continuously fluctuate.
Under-loaded instances usually require fewer resources and are less able to utilize the resources, resulting in resource wastage of part \red{\ding{173}} in Figure~\ref{fig:util-motiv}.

%To address the above issues, serverless systems can attempt from two fronts.
To mitigate the two issues, nowadays serverless systems can leverage \emph{overcommitment}~\cite{10.1145/3542929.3563470,10.1145/3620678.3624645} and \emph{autoscaling}~\cite{288784}.
%can attempt from two fronts.
First, for part \blue{\ding{172}} in Figure~\ref{fig:util-motiv}, overcommitment is an intuitive solution, i.e., deploy more instances on a server.
This approach could be implemented by the scheduler, as the scheduler is responsible for deciding how instances are deployed.
Second, for part \red{\ding{173}}, it could be helpful to reduce the load fluctuations within instances, trying to keep the load close to the saturated load for each instance.
This can be achieved by the autoscaler, which is responsible for observing load fluctuations and adjusting the number of instances accordingly.

Nevertheless, it is challenging to apply the two approaches while achieving effectiveness with practical cost.
%The challenges are as follows:
First, since overcommitment may lead to performance degradation and QoS violation, it is a common practice to predict performance before scheduling an instance to avoid QoS violation.
%it is best for the scheduler to predict performances when scheduling, and increase the instance deployment density as much as possible without violating QoS.
%However, it is challenging to \emph{accurately predict instances' performance with a practical cost, even if their load is assumed to be stable at the saturated value.}
However, it is challenging to \emph{accurately predict instances' performance with a practical cost.}
Although prior predictor-based schedulers have made great progress~\cite{10.1145/3542929.3563470,10.1145/3458817.3476215,10.1145/2451116.2451125,10.1145/2155620.2155650,8005341,10.1145/3274808.3274820,10.1145/2485922.2485975}, they fall short in achieving both high accuracy and low cost, since the two goals can be a trade-off:
making accurate prediction requires considering complex \emph{interference} on multiple resources imposed by highly heterogeneous colocated functions, but the complex computation inevitably introduces high overheads.
For example, some prior works predict with specially designed machine learning models~\cite{10.1145/3458817.3476215}, which is costly for scheduling.
%Since cold start latency mainly consists of two main components: scheduling latency and instance initialization latency, with instance initialization being optimized by a lot of work to even $<$1ms (Table~\ref{t:sys-startup}),
%scheduling with model inference ($\sim$20ms) could possibly become new bottleneck~\cite{10.1145/3620678.3624785}.
Since cold start latency mainly consists of scheduling latency and instance initialization latency, with initialization being optimized by a lot of work to even $<$1ms,
scheduling with inference ($\sim$20ms or higher) could become a new bottleneck~\cite{10.1145/3620678.3624785}.
To compromise, nowadays public clouds still use heuristic policies for scheduling, thus failing to accurately predict performance and providing limited improvement in resource utilization.

Second, to efficiently utilize resources under load fluctuation, the autoscaler needs to quickly respond to the load variation, evict instances and free the resources when the load drops. 
Intuitively, the faster autoscaling can react to load fluctuations, the more efficiently resources can be used. 
However, \emph{more sensitive and frequent scaling also means more additional cold starts, resulting in the trade-off between resource utilization and cold start overheads.}
It is challenging to achieve both goals simultaneously.

\iffalse irregular fluctuation
To make matters worse, it is almost impossible to predict such load variations.
According to Azure trace and related analysis based on it~\cite{fuerst2021faascache,shahrad2020serverless,10.1145/3503222.3507750,10.1145/3567955.3567960}, although the load fluctuation can have some patterns (e.g., diurnal patterns) on a relatively large time scale, it can not be accurately predicted in a short time period.
Therefore, the more sensitive the instance scaling is and the more frequent cold starts, the more difficult it is for the instance pre-warm approach to effectively address the issue.
\fi

We present two key insights to tackle the aforementioned challenges.
First, the trade-off between scheduling costs and efficiency is due to the fact that these complex decisions need to be made at the time of scheduling, which involves costly predictions.
Therefore, it can be beneficial if we \emph{decouple prediction and decision making}.
Specifically, we can predict in advance the performance of possible incoming instances and save the results.
If the incoming instance matches the advance predicted scenario, the scheduling can be made by directly checking the prepared decisions without model inference. 
This provides a scheduling fast path. 
%Inspired by serverless's high scalability nature, we can make predictions for the next instances of the existing functions. 
%This can help ensure most schedulings go through the fast path.

Second, the trade-off between resource utilization and cold starts arises from that instances do not release resources until they are evicted.
Instead, we can \emph{decouple resource releasing and instance eviction.}
%\highlight{
%Specifically, operations like sending requests to fewer instances could achieve similar effect of releasing resources as eviction.
%Releasing/reclaiming resources without instance eviction/creation can be less costly, thus tolerating higher sensitivity for better resource utilization.
Specifically, operations like sending requests to fewer instances could achieve similar effect of releasing resources as eviction.
Releasing resources without instance eviction can be less costly, and thus can be applied with higher sensitivity for better resource utilization.
Inspired by the two insights, we propose \sys, a serverless system that harmonizes efficiency and practicability in improving resource utilization, tackling the challenges with two techniques.
First, \sys achieves accurate and low cost performance prediction with \emph{pre-decision scheduling}, which decouples prediction and decision making, providing a scheduling \emph{fast path} without inference.
For every deployed function on a server, \sys predicts in advance their \emph{capacities} on that server with a model.
A function's capacity means the maximum number of its instances can be deployed on the server under the interference of current neighbors without violating everyone's QoS.
To schedule a new instance of a deployed function, \sys only needs to compare the function's capacity with the number of its instances to determine whether the scheduling can be successful or not.
%When scheduling a new instance of a deployed function to that server, \sys only needs to compare the capacity with the function's concurrency to determine whether the scheduling can be successful or not.
% Moreover, to determine whether deploying an instance will cause other colocated functions to violate their QoS,
% \sys further designs an \emph{asynchronous update} approach, ensuring such model-inference-dependent validation process can be performed outside the scheduling critical path.
Moreover, \sys designs an \emph{asynchronous update} approach, which keeps the capacities up-to-date without incurring model inference to the scheduling critical path,
and applies a \emph{concurrency-aware scheduling} to batch the scheduling of concurrent incoming instances when load spikes occur.
Results with real-world patterns show that >80\% scheduling goes through the fast path.
%Our evaluation shows that most (>80\%) scheduling goes through such fast path.
%Only when an instance is newly scheduled to a node where the capacity is not ready, it needs to go through the \emph{slow path} and uses the prediction model to prepare the capacity value when scheduling.

%Second, in replace of the request router's load balancing convention, \sys designs a load consolidation approach that cooperates with the instance scheduler.  
Second, \sys designs a \emph{dual-staged scaling} method.
When load drops, before instances are evicted, \sys first adjusts the routing with a higher sensitivity, sending requests to fewer instances, thus releasing some of the instances' resources.
Then, if the load rises again before idle instances are evicted, \sys can get the instances working again by re-routing with minimum overhead.
Third, to cope with the case where a node is full and causes idle instances on it can not be converted to the saturated state,
\sys migrates idle instances to other nodes in advance to hide the overhead of the required cold starts.  

We present a prototype of \sys based on an open-source serverless system, OpenFaaS~\cite{openfaas}.
We evaluate it using ServerlessBench~\cite{socc20-serverlessbench}, FunctionBench~\cite{kim2019practical}, and applications with real-world traces from \hwcloud.
The results show that on real-world traces, compared to a state-of-the-art model-based serverless scheduler~\cite{10.1145/3458817.3476215}, \sys incurs 81.0\%--93.7\% lower scheduling costs and 57.4\%--69.3\% lower cold start latency with \textit{cfork}~\cite{10.1145/3503222.3507732}.
It also achieves the highest resource utilization compared to all baseline schedulers (54.8\% higher instance density than Kubernetes).

\section{Motivation}
\label{s:motiv}

\subsection{Background and Motivation}

%% 1. function v.s. instance
%% 3. QoS: tail latency. + others
%% 2. LB, router
%% 4. Scheduler: responsibility, policy
%% 5. Autoscaler and autoscaling:
%%     5.1 Saturated value
%%     5.2 Keep-alive: presents sensitivity.

\myparagraph{Notations and terms.}
% Serverless \underline{function} represents the basic scheduling unit of an application in serverless platform.
% A function can have multiple \underline{instances} serving requests, each allocated with a specified amount of resources by user configurations.
% We use \underline{Quality-of-Service (QoS)} to represent the performance objectives of functions.
% In our platform, QoS is defined with tail latency, and it can be extended to other metrics.
The term \underline{function} denotes the basic scheduling unit within a serverless platform.
A function might comprise multiple \underline{instances} to serve requests.
Each instance is allocated with specified resources by user configurations.
\underline{Quality-of-Service (QoS)} is utilized to describe the performance targets for functions.
In our platform, QoS is primarily characterized by tail latency; however, the definition can be extended to other metrics.

% Moreover, serverless platforms are expected to guarantee the performance objectives of functions, which are usually quantified as their Quality-of-Service (QoS).
% In practice, the QoS of a function is usually defined in terms of the tail latency not exceeding a configured threshold.

% \myparagraph{Key components of serverless platforms responsible for loads and utilization.}
% A serverless platform includes \underline{router}, \underline{autoscaler}, and \underline{scheduler}, to manage loads and instances.
% First, router dispatch user requests to instances of a function, usually using \emph{load balance} policy, which equally distribute requests to instances.
% Second, autoscaler~\cite{knative-scalar,openfaas-scalar} is responsible to determine the number of instances.
% It will use a pre-defined threshold RPS (which we called \underline{saturated load}) to determine whether it is necessary to create or evict instances accordingly,
% to avoid instance overloaded or resource waste.
% Different from creating instances, the autoscaler evicts instances with a keep-alive approach, which does not immediately evict instances when the expected number of instances drops, but waits for a ``keep-alive duration''.
% The keep-alive duration can reflect the sensitivity of the autoscaler. 
% Shorter keep-alive duration means higher sensitivity and more frequent evictions.
% Last, when a new instance is created, a serverless \emph{scheduler} is responsible for assigning a specific node for it to deploy the instance, which will consider several factors like the resource requirements.

\begin{figure}[t]
  \centering
   \begin{minipage}[t]{0.98\linewidth}
    \centering
    \includegraphics[width=0.78\textwidth]{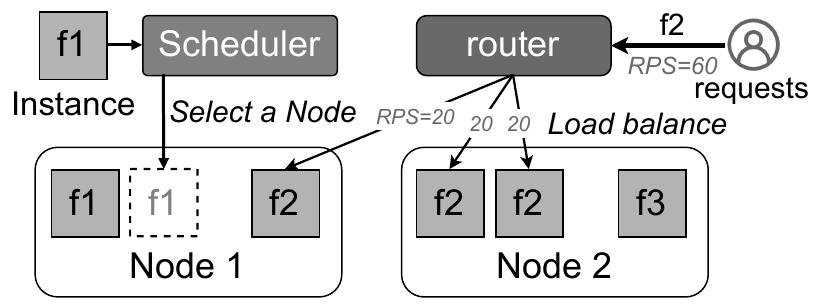}
    \footnotesize
    \end{minipage}
	\caption{\textbf{Schedulers and routers.}}
%\MZY{The x and y axis should use scheduling costs and Instance QoS.}\DD{fixed.}
  \label{fig:back-preliminary-concept}
\end{figure}

\myparagraph{Components for load and utilization management in serverless platforms.}
First, \underline{router} dispatches user requests to available instances of a function, typically employing a \emph{load balancing} strategy to ensure equal distribution of requests among instances (Figure~\ref{fig:back-preliminary-concept}).
Second, \underline{autoscaler}~\cite{knative-scalar,openfaas-scalar} is responsible for determining the number of instances.
It uses a pre-determined RPS threshold, referred to as the \underline{saturated load}, to determine whether it is necessary to create or evict instances, to avoid instance overloaded or resource waste.
Different from creating instances, the autoscaler evicts instances with a \underline{keep-alive} approach, which does not immediately evict instances when the expected number of instances drops, but waits for a ``keep-alive duration''.
The keep-alive duration can reflect the sensitivity of the autoscaler --- shorter duration means higher sensitivity and more frequent evictions.
Last, when a new instance is created, a \underline{scheduler} is responsible for assigning a specific node for it to deploy the instance, which will consider several factors like the resource requirements.

\begin{figure}[t]
  \centering
 \setlength{\belowcaptionskip}{-2pt}
  \setlength{\abovecaptionskip}{-1pt}
   \begin{minipage}[t]{0.98\linewidth}
    \centering
    \includegraphics[width=0.92\textwidth]{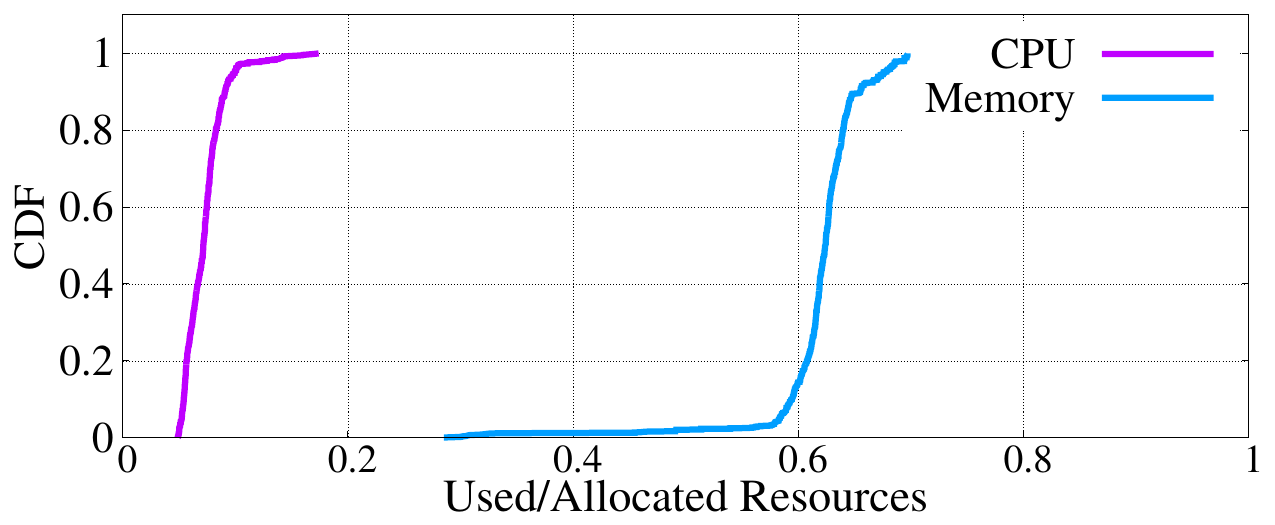}
    \footnotesize
    \end{minipage}
	\caption{\textbf{Statistics on the ratio of actual resource usage to allocated resources.} The data was collected on \hwcloud over one month from a region. About 80\% of servers only use 8\% CPU and 64\% memory resources.}
%\MZY{The x and y axis should use scheduling costs and Instance QoS.}\DD{fixed.}
  \label{fig:motiv-util-cdf}
\end{figure}

%\myparagraph{Serverless background.}
%\myparagraph{Resource utilization of serverless functions.}
\myparagraph{Low resource utilization in serverless platforms.}
%\myparagraph{Low resource utilization in serverless platforms.}
%\highlight{
Optimizing resource usage is one of the top cloud initiatives~\cite{flexera-2023}, as every 1\% increase in utilization could mean more than millions of dollars a year in cost savings.
However, production traces show that the resources of serverless platforms are still severely under-utilized (Figure~\ref{fig:motiv-util-cdf}).
In general, the wastage mainly arises from two parts: part \blue{\ding{172}} and \red{\ding{173}} in Figure~\ref{fig:util-motiv}.
%}

\begin{figure}[t]
  \centering
  \includegraphics[width=0.38\textwidth]{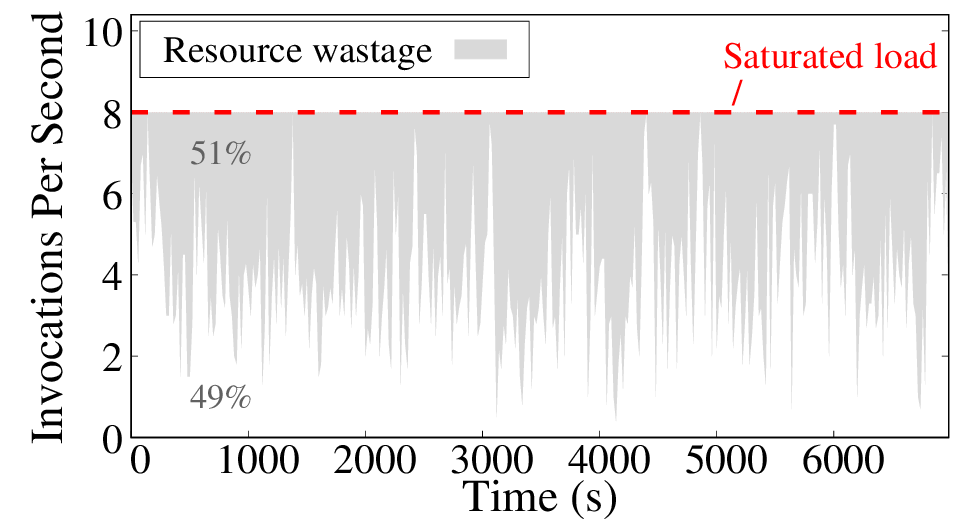}
	\caption{\textbf{Fluctuation of user load per instance.}}
%\MZY{The x and y axis should use scheduling costs and Instance QoS.}\DD{fixed.}
  \label{fig:back-invocation-fluctuate}
\end{figure}

%\highlight{
Part \blue{\ding{172}} is because the resources allocated by the user are usually conservative, considering the worst cases that suffer the severest resource interference.
Therefore, the allocated resources could be even higher than the actual demand of instances that are processing saturated load.
Part \red{\ding{173}} is because the load served by each instance constantly fluctuates, and under-loaded instances are less able to fully utilize the resources.
For example, Figure~\ref{fig:back-invocation-fluctuate} shows the average RPS served by an instance for one of the most popular functions in the trace of \hwcloud.
If the instance is always considered saturated, \textbf{51\%} of resources could possibly be wasted.
%}

\myparagraph{Opportunities for improving resource utilization.}
%\highlight{
To mitigate the issues, serverless systems could adopt two approaches respectively.
First is overcommitment for part \blue{\ding{172}}, i.e., the scheduler could deploy more instances on a server.
Second is autoscaling for part \red{\ding{173}}, i.e., the autoscaler dynamically evicts instances when the load drops, reducing the overestimation of resource demands for under-loaded instances.
Applying the two approaches, our objective is to simultaneously achieve \emph{effectiveness} in improving resource utilization with \emph{practical cost}.
%}

\begin{table*}[t]
    \caption{\textbf{A comparison between \sys and previously developed methods.} 
    Scalable profiling requires no more than O($n$) complexity, where $n$ represents the number of functions, 
    $k$ denotes to the number of instances that can be deployed on a server.
    Pythia, due to have a model for each function, requires O($n^2$) complexity. 
    Whare-map requires O($n^k$) complexity to profile all colocation combinations where $k$ denotes to the number of co-runners allowed on a server. 
    Owl profiles the combination of two functions with $k$ instances, requiring O($n^{2}k$) complexity.
    \textit{``Fast scheduling''} implies that the scheduler makes an decision within a few milliseconds ($\sim$1ms). Gsight requires more than 20ms for scheduling (\textsection\ref{subs:eval-perf}). 
    }
    \label{tab:prior-works}
    \footnotesize
    \begin{tabular}{l|l|l|llll|l}
        \toprule
        \multirow{3}{*}{\textbf{Workload}}           & \multirow{3}{*}{\textbf{System}} & \multirow{3}{*}{\textbf{Year}} & \multicolumn{4}{c|}{\textbf{Efficient QoS-Aware Scheduler with Low Cost?}}                                                                                                                                   & \multirow{3}{*}{\textbf{\tabincell{l}{Handling Load\\Dynamicity?}}} \\ \cline{4-7}
                                    &                         &                       & \multicolumn{1}{l|}{\multirow{2}{*}{\tabincell{l}{\textbf{Prediction model?}}}} & \multicolumn{1}{l|}{\multirow{2}{*}{\textbf{\tabincell{l}{Accurate\\Prediction?}}}} & \multicolumn{2}{c|}{\textbf{Overhead?}}                    &                                            \\ \cline{6-7}
                                    &                         &                       & \multicolumn{1}{l|}{}                                                   & \multicolumn{1}{l|}{}                                      & \textbf{Profiling cost?}         & \textbf{Fast scheduling?}        &                                            \\ \hline
        \multirow{3}{*}{Serverless}         & \textbf{Jiagu}      &                            & \multicolumn{1}{l|}{\textbf{Global statistical model}}                 & \multicolumn{1}{l|}{\green{\cmark}}                  & \green{O($n$)}                & \green{\cmark}                 & \green{\textbf{Dual-staged scaling}}                          \\
                                            & Owl~\cite{10.1145/3542929.3563470} & 2022                                    & \multicolumn{1}{l|}{Historical information}                   & \multicolumn{1}{l|}{\red{Limited}}                                          & \red{O($n^{2}k$)}                 & \green{\cmark}                 & Autoscaling                                               \\
                                            & Gsight~\cite{10.1145/3458817.3476215} & 2021                                  & \multicolumn{1}{l|}{Global statistical model}                 & \multicolumn{1}{l|}{\green{\cmark}}                  & \green{O($n$)}                  & \red{\xmark}                                        &  Autoscaling                                              \\ \hline
        \multirow{4}{*}{\tabincell{c}{Monolithic\\Service}} & Pythia~\cite{10.1145/3274808.3274820} & 2018                                    & \multicolumn{1}{l|}{\tabincell{l}{Per-function Linear model}} & \multicolumn{1}{l|}{\green{\cmark}}                  & \red{O($n^{2}$)}                                          & \green{\cmark}                 &    Autoscaling                                            \\
                                            & Paragon~\cite{10.1145/2451116.2451125} & 2013                                 & \multicolumn{1}{l|}{Heuristic}                                & \multicolumn{1}{l|}{\red{NA}}                                         & \green{O($n$)}                  & \green{\cmark}                 &  Autoscaling                                              \\
                                            & Whare-map~\cite{10.1145/2485922.2485975} & 2013                           & \multicolumn{1}{l|}{Historical information}                   & \multicolumn{1}{l|}{\red{Limited}}                  & \red{O($n^{k}$)}                                          & \green{\cmark}                 &   Autoscaling                                             \\
                                            & Bubble-up~\cite{10.1145/2155620.2155650} & 2011                              & \multicolumn{1}{l|}{Heuristic}                                & \multicolumn{1}{l|}{\red{Inaccurate}}                                         & \green{O($n$)}                  & \green{\cmark}                 &   Autoscaling                                            \\ \hline
	    {\tabincell{c}{Resource\\Schedulers}}         &  \multicolumn{3}{c|}{\tabincell{c}{Aquatope~\cite{10.1145/3567955.3567960}, Cilantro~\cite{288542}, Sinan~\cite{zhang2021sinan}\\FIRM~\cite{10.5555/3488766.3488812}, Parties~\cite{10.1145/3297858.3304005}, Heracles~\cite{10.1145/2749469.2749475}, ...}}      & \multicolumn{4}{c}{Resilient to runtime variations with limitations (\textsection\ref{s:relwk}).}                 \\
        \bottomrule
        \end{tabular}

\end{table*}
% \usepackage{multirow}

\subsection{Challenges}
%\highlight{
However, achieving both efficiency and practicability can be challenging, since the two goals usually have trade-offs.
%Specifically, the challenges of achieving both goals for the scheduler and the autoscaler are as follows.
%}

\subsubsection{Challenges for the Scheduler}
\label{s:requirement1}
%\highlight{
Although overcommitment helps improve resource utilization, it could also cause performance degradation, possibly resulting in QoS violation. %of overcommitted instances.
Therefore, a common approach for the scheduler is to predict instances' performances when scheduling, maximizing instance deployment density while not violating instances' QoS.
%The goal of reaching both practicability and efficiency here means that, the scheduler should \textit{simultaneously} achieve: accurate QoS violation prediction (with scalable profiling overhead) and low scheduling latency.
The goal means that the scheduler should \textit{simultaneously} achieve: accurate QoS violation prediction (with scalable profiling overhead) and low scheduling latency.
Although prior schedulers have made great progress for the serverless scenario~\cite{10.1145/3542929.3563470,10.1145/3458817.3476215} compared with schedulers~\cite{10.1145/2451116.2451125,10.1145/2155620.2155650,8005341,10.1145/3274808.3274820,10.1145/2485922.2485975} designed for monolithic services, they fall short of achieving both goals.
Table~\ref{tab:prior-works} summarizes the key differences of prior schedulers against our objectives.
%}

\iffalse
% \begin{figure}[t]
%   \centering
% %  \setlength{\belowcaptionskip}{-2pt}
%   % \setlength{\abovecaptionskip}{-1pt}
%    \begin{minipage}[t]{0.98\linewidth}
%     \centering
%     \includegraphics[width=0.85\textwidth]{figs/ServerlessPrior.pdf}
%     \footnotesize
%     \end{minipage}
% 	\caption{\textbf{A comparison between \sys and previous serverless schedulers}. \sys is the first serverless QoS-aware scheduler to achieve both accurate prediction and low cost.}
% %\MZY{The x and y axis should use scheduling costs and Instance QoS.}\DD{fixed.}
%   \label{fig:motiv-serverless-prior}
% \end{figure}
\fi

\myparagraph{Accurate QoS prediction.}
The key to accurately predicting the QoS violation is to quantify the resource interference by colocated function instances, which is challenging.
First, in the serverless scenario, the number of colocation combinations grows exponentially with the number of functions, and ultimately becomes practically infinite.
Therefore, the profiling overhead should be scalable, otherwise it could be impossible to profile such a large number of colocation combinations.
% Consequently, prediction methods with unscalable profiling overhead, for example, exhaustively record the historical performance information for prediction, is impractical.
%Secondly, resource interference can be complex, while plentiful and heterogenenous functions with arbitrary code exerting various pressure on multiple resources.
Second, resource interference can be complex, while plentiful and heterogeneous functions exert various pressures on multiple resources.
The large number of colocation combinations further amplifies the interference complexity.
% Therefore, predicting with these complex and rich features, the model requires computationally intensive calculations.

%\highlight{
Some prior schedulers do not meet the goal yet. 
For example, some schedulers predict based on historical performance information~\cite{10.1145/3542929.3563470,10.1145/2485922.2485975}, so that their predictive ability is limited and can only predict the performance of profiled colocated instance combinations.
Their profiling cost is also unscalable (shown in Table~\ref{tab:prior-works}).
Schedulers that train a model for each function has unacceptable profiling and training overhead~\cite{10.1145/3274808.3274820}.
Schedulers using heuristic algorithms are limited in their ability to accurately predict performance due to overly simplistic models~\cite{10.1145/2155620.2155650}, or does not actually predict performance at all~\cite{10.1145/2451116.2451125}.
\begin{table}[t]
    \centering
	\footnotesize
    \caption{\textbf{State-of-the-art serverless systems with startup optimizations.} ``Container startup'' presents the runtime startup latency reported by public reports or papers. ``Scheduling overhead'' presents the overhead of scheduling in a cold startup with model-based methods. We use 21.78ms as the model-based scheduling costs, the average result of our ported Gsight~\cite{10.1145/3458817.3476215} model.
    %As shown in the table, scheduling costs are non-trivial even for commercial systems like AWS Lambda, and can be dominant for state-of-the-art systems.
    }
    \label{t:sys-startup}
    \begin{tabular}{lcc}
    \hline
	    \textbf{System}                                                        & \textbf{Container startup} & \textbf{Scheduling overhead}  \\ \midrule
    AWS Snapstart~\cite{aws-snapstart}                                                                         & $\sim$100ms                       & $\sim$21.8\% \\ %\hline
    Replayable~\cite{wang2019replayable}            & 54ms                              & 40.3\%     \\ %\hline
    Fireworks~\cite{10.1145/3492321.3519581} & $\sim$50ms                        & $\sim$43.6\% \\ %\hline
    SOCK~\cite{oakes2018sock}                           & 20ms                                       & 108.9\%     \\ %\hline
    Molecule~\cite{10.1145/3503222.3507732}                                              & 8.4ms                                   & 2.6x    \\ %\hline
    SEUSS~\cite{cadden2020seuss}                                          & 7.5ms                                    & 2.9x    \\ %\hline
    Catalyzer~\cite{10.1145/3373376.3378512} & 0.97ms                                  & 22.5x   \\ %\hline
    Faasm~\cite{shillaker2020faasm}                     & 0.5ms                                     & 43.6x   \\ \hline
    \end{tabular}
    \end{table}

% \usepackage{booktabs}
% \usepackage{multirow}

\iffalse
   Molecule~\cite{10.1145/3503222.3507732}                                              & 8.4ms                                   & 259.3\%    \\ %\hline
    SEUSS~\cite{cadden2020seuss}                                          & 7.5ms                                    & 290.4\%    \\ %\hline
    Catalyzer~\cite{10.1145/3373376.3378512} & 0.97ms                                  & 2245.4\%   \\ %\hline
    Faasm~\cite{shillaker2020faasm}                     & 0.5ms                                     & 4356\%   \\ \hline

\fi

\myparagraph{Low scheduling latency.}
The scheduling process is involved in every instance's cold start.
%It is an important but often ignored part of the cold start latency.
Given the recent advancements in optimizing the instance initialization cost to a few milliseconds or even sub-milliseconds~\cite{DBLP:conf/usenix/WangCTWYLDC21, akkus2018sand, wang2019replayable,10.1145/3373376.3378512, oakes2018sock, jianightcore, 10.1145/3492321.3519581, 10.1145/3503222.3507732, cadden2020seuss, shillaker2020faasm}, it becomes imperative to ensure that the scheduling cost remains low as well.
Otherwise, the scheduling cost can become the new bottleneck in the startup cost~\cite{10.1145/3620678.3624785}.
As shown in Table~\ref{t:sys-startup}, we use the average scheduling costs of Gsight~\cite{10.1145/3458817.3476215} (our ported version) to illustrate the overhead of scheduling over the total cold startup costs in these systems.
Although it can accurately predict the performance and consider serverless-specific features like partial interference, it can incur non-trivial overhead even in commercial systems like AWS Lambda with Snapstart~\cite{aws-snapstart} (>20\%).
Scheduling cost can be dominant for cold starts with state-of-the-art optimizations like Catalyzer~\cite{10.1145/3373376.3378512} and Faasm~\cite{shillaker2020faasm}.
In short, schedulers need to make decisions within a few milliseconds to avoid hindering the startup process.

%\subsubsection{Maximizing Resource Utilization under Load Fluctuation}
%\subsubsection{Challenges of Reducing Resource Wastage Caused by Load Fluctuation}
\subsubsection{Challenges for Autoscaling}
%\highlight{
The autoscaling feature commonly helps mitigate the defect.
When load drops, the autoscaler would dynamically evict instances, thus freeing space for deploying new instances, preventing those under-loaded instances from wasting resources (typically every instance can be under-loaded because of load balancing).
Intuitively, a more sensitive autoscaler could better utilize resources in response to load fluctuation.
However, more frequent scaling would also result in more additional cold starts.
It means the \emph{trade-off between resource utilization and cold start costs.}
%}

%\highlight{
The irregularity of load fluctuations further complicates the problem.
Prior works have analyzed production traces and tried to predict the invocation patterns~\cite{fuerst2021faascache,shahrad2020serverless,10.1145/3503222.3507750,10.1145/3567955.3567960}, pre-warming instances accordingly to reduce cold starts.
However, user loads exhibit only moderate regularity over extended periods of time (e.g., diurnal patterns), but are extremely unpredictable over shorter intervals.
For example, the average coefficient of variable (CV) over the number of requests in a minute could be more than 10~\cite{285173} in the Azure trace~\cite{shahrad2020serverless}.
The more frequent scaling also means a finer-grained pre-warming prediction and a worse prediction accuracy.

\subsection{Insights}
\label{s:traces}
We observe two insights that help to tackle the two challenges.

\myparagraph{Insight-1: Prediction and decision making can be decoupled.}
The trade-off between prediction accuracy and scheduling cost arises from the coupling of the prediction and decision making, i.e., the costly model inference is usually made during scheduling.
We can \emph{decouple prediction and decision making}.
Specifically, the scheduler can assume the interference environment before the new instance arrives, and make advance predictions based on the assumptions.
When a new instance arrives and the interference environment at the time of scheduling matches our previous assumption, the scheduling can be made by directly checking the prepared decisions.
This provides a scheduling fast path without inference.

\iffalse old insight of capacity
% First, rather than predicting the QoS violation of single instances, the scheduler can first predict the \emph{capacity}, i.e., how many concurrent instances of the same function can be safely deployed to a node without violating QoS.
% It means when the number of currently running instances is below the capacity, we can directly schedule a new instance of the same function without prediction (i.e., fast path), significantly reducing the scheduling costs.
% Serverless's \concurrency nature implies that most scheduling processes will take the fast path.
% % Only when an instance is newly deployed to a server where its capacity is not calculated, the scheduler will predict the capacity with the model before deployment (i.e., the slow path).
% In short, the introduction of \textit{capacity} can meet the first requirement: it can significantly reduce the cost of model inference when scheduling, while still taking advantage of the accurate prediction of the machine learning model.
\fi

\myparagraph{Insight-2: Resource releasing and instance eviction can be decoupled.}
%\highlight{
%Second, 
The trade-off between resource utilization and cold start overheads arises from the coupling of resource releasing and instance eviction.
%\highlight{
Instead, we can \emph{decouple resource releasing and instance eviction.}
Specifically, even if an instance is not evicted, we could adjust the routing and do not send requests to it.
It consolidates the loads of under-loaded instances to fewer instances to reduce the waste of resources caused by treating under-loaded instances as saturated instances, achieving a similar resource releasing effect as an actual eviction.
The overhead of adjusting the routing is much smaller than actual cold start overheads.
%}
Therefore, in this way, we could release/reclaim resources with higher sensitivity to cope with load fluctuations, while avoiding excessive additional cold start overheads.

\section{Design Overview}
\label{s:design}

%This section decribes the design of \sys, a QoS-and-concurrency-aware serverless scheduler.
%for serverless platforms.

\begin{figure}[t]
    \centering
    \includegraphics[width=0.48\textwidth]{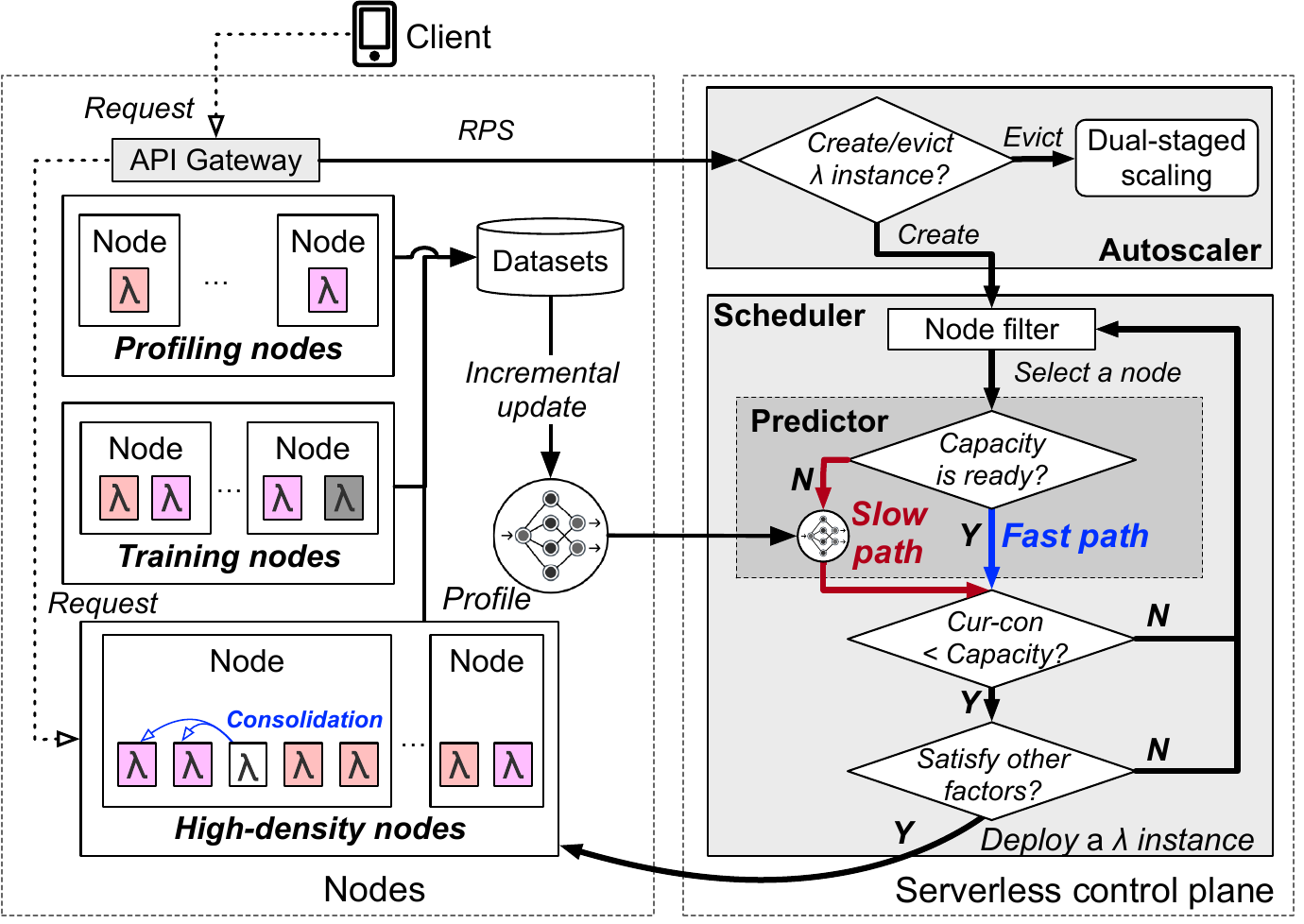}
    \caption{\textbf{\sys overview.} 
    % \sys is a serverless scheduler that utilizes a concurrency-aware ML model to predict the capacity of functions. The ML model is trained offline and can be incrementally updated during runtime. \sys can schedule an instance immediately (no inference) when capacity is ready (i.e., fast path) and update the capacity using the model (i.e., slow path) when necessary.
    }
    \label{fig:design-overview}
\end{figure}

We propose \sys, an efficient and practical QoS-aware serverless system that tackles the two challenges of improving resource utilization with the insights.
The overall design is shown in Figure~\ref{fig:design-overview}.
First, \sys designs a pre-decision scheduling that could make accurate predictions with low scheduling latency (\textsection\ref{s:scheduler}). 
When an instance is created, \sys predicts its performance after deployment, trying to increase the instance deployment density without violating QoS to fully utilize the resources.
Second, \sys adopts a dual-staged scaling design that efficiently utilizes resources under load fluctuation with minimum overhead (\textsection\ref{s:load-consolidation}).

%\myparagraph{Node partitioning in clusters.}
\myparagraph{Cluster setting.}
To apply \sys, our cluster is partitioned into three types of nodes. 
A small proportion of nodes are profiling nodes and training nodes, deploying instances whose runtime information is collected to construct function profiles and the dataset to train the model used by \sys's scheduler.
\sys schedules instances that actually handle user requests to high-density nodes, maximizing the resource utilization of these nodes.
%\highlight{
	The training, profiling and high-density nodes are homogeneous.
%}

%Our prototype on OpenFaaS has 4.5k+ LoCs changes.
%\sys is also developed and discussed at \hwcloud, and we will present several lessons learned from industry development (\textsection\ref{subs:eval-industry}).
\myparagraph{User configurations and QoS.}
% Defining QoS is important to ensure user experience, which is essential to attract and retain users.
%\highlight{
In our platform, when uploading a function, a user is expected to specify the allocated resources (CPU and memory) and the saturated load of the function (optional with default value).
The QoS requirement of a function is established by the provider according to the function's previously monitored performances, as an important reference metric for overcommitment.
Moreover, in \hwcloud, we can establish specific QoS agreements with top-tier customers, since their workload could be the majority in the cloud.
%For retail users, a range of measures is planned to support their QoS in the future.
%}

\section{Pre-decision Scheduling} % with the Model
\label{s:scheduler}

\subsection{\sys's Prediciton Model}

\myparagraph{Prediction model.}
\sys's ability of QoS-aware scheduling requires accurate prediction of the functions' performance under resource interference.
The prediction model is based on Random Forest Regression (RFR). 
Specifically, \sys's regression model has the following form:
$$
P_{A\cup\{B,C,...\}} = RFR\{P_A, R_A, C_A, R_B, C_B, R_C, C_C,...\}
$$

where, $P_{A\cup\{B,C,...\}}$ is function $A$'s performance\footnote{
    Although we use P90 tail latency as our performance metrics, it is a general model that can be extended to other metrics like costs.}
under the interference of functions $B$ and $C$. 
$R_A$, $R_B$ and $R_C$ is the profile matrix of function $A$, $B$ and $C$ respectively.
\sys uses a single instance's mutliple resource's utilization as its profile (Table~\ref{tab:metrics}), and adopts a \textit{solo-run} approach to collect the profile metrics of functions (details in \textsection\ref{s:impl}).
$C_A$, $C_B$ and $C_C$ are concurrency information of the three functions.
The concurrency information includes two parts: the number of saturated instances and the number of \cached instances (detailed in \textsection\ref{s:load-consolidation}). 
$P_A$ is the solo-run performance of function $A$.
% We choose Random Forest Regression as the prediction model.

%Since different instances of the same function have the same profile, compared with Gsight's model, we merge the features of the same function's instances and introduce \textit{concurrency} as a new feature for a function.
The model is proven effective in our public cloud, and also utilized in prior systems~\cite{10.1145/3458817.3476215,10.1145/3620678.3624645}.
Moreover, compared to prior models that predict the performance in the instance granularity~\cite{10.1145/3458817.3476215,10.1145/3274808.3274820}, since different instances of the same function are homogeneous and performs similarly on the same server, the model predicts the performance in function-granularity.
We merge the features of the same function's instances and introduce \textit{concurrency} as a new feature for a function.
This effectively reduces the input dimensions, resulting in less training overhead and possible mitigation of the ``curse of dimensionality''~\cite{bellman2003dynamic}.
\sys is also flexible to utilize other prediction models.

% \usepackage{booktabs}
% \usepackage{multirow}

\begin{table}[t]
	\setlength{\belowcaptionskip}{0pt}
    \setlength{\abovecaptionskip}{0pt}
    \centering
    \footnotesize
    \caption{\textbf{Profiling metrics.}}
    \label{tab:metrics}
    \begin{tabular}{@{}lp{0.28\textwidth}@{}}
    \toprule
    \textbf{Metrics}        &  \textbf{Description}         \\ \midrule
    mCPU       &  CPU utilization      \\
    Instructions            &  Instructions retired     \\
    IPC                     &  Instructions per cycle     \\
    Context switches        &  Switching between privilege modes    \\
    MLP & Efficiency of concurrent access indicated by Memory Level Parallelism (MLP) \\
    L1d/L1i/L2/LLC MPKI     & Cache locality indicated by misses per thousand instructions (MPKI)  \\
    TLB data/inst. MPKI     & TLB locality indicated by MPKI    \\
    Branch MPKI             & Branch predictor locality indicated by MPKI     \\
    Memory bandwidth        &  Memory usage and performance     \\ \bottomrule
    \end{tabular} \\
\end{table}

\myparagraph{Inference overhead.}
% Each input of \sys's model describes a specific colocation environment, and the model predicts the performance of a function amid this interference.
In practice, a policy may require predicting different functions' performances under various colocation environments, and the colocation environments can be described by multiple inputs.
Notably, learning frameworks (e.g., sklearn\cite{sk-learn}) have optimizations to infer multiple simultaneous inputs, which incurs trivial additional cost (\textsection\ref{s:eval}).
Therefore, in the descriptions that follow, if multiple predictions can be merged in the form of multiple inputs, we refer to such inference overhead as ``once'' inference overhead.

\subsection{Capacity and Capacity Table}
\label{s:capacity-and-delayed-update}
\iffalse old start
% With the ability of predicting performance, the most intuitive way of scheduling is that, when scheduling every instance, predict whether the instance will violate QoS when deployed on a server, thus determining whether it can be scheduled on that server, as shown in Figure~\ref{fig:capacity-batch}(a).
% %However, incurring a model inference at every scheduling process can be costly (\textsection\ref{s:requirement1}).
% %Therefore, \sys adopts further optimizations to hide the overhead.
% However, incurring a model inference at every scheduling process can be costly.
% \sys adopts novel techniques to mitigate the costs.
\fi
Inspired by the first insight, to accurately predict instances' performances without incurring excessive cost to scheduling, we can decouple prediction and decision making.
However, making predictions requires knowledge of what the incoming instance is and what the interference environment on the server is, which is not known until scheduling (decision making).
Considering there are infinite number of functions, it is also impossible to traverse all possible interference environments. 
Therefore, the decoupling can be challenging.

\begin{figure}[t]
  % \setlength{\belowcaptionskip}{-8pt}
  %  \setlength{\abovecaptionskip}{0.5pt}
% 
  %  \setlength{\abovecaptionskip}{-5pt}
  % \begin{minipage}[t]{0.33\linewidth}
  %   \centering 
  %   \includegraphics[scale=0.40]{figs-eval/trace-concurrent-cdf.pdf}
  %   \footnotesize
  %   \textbf{(a) Function distribution.}
  %   \end{minipage}
  \begin{minipage}[t]{0.49\linewidth}
    \centering 
    \includegraphics[scale=0.30]{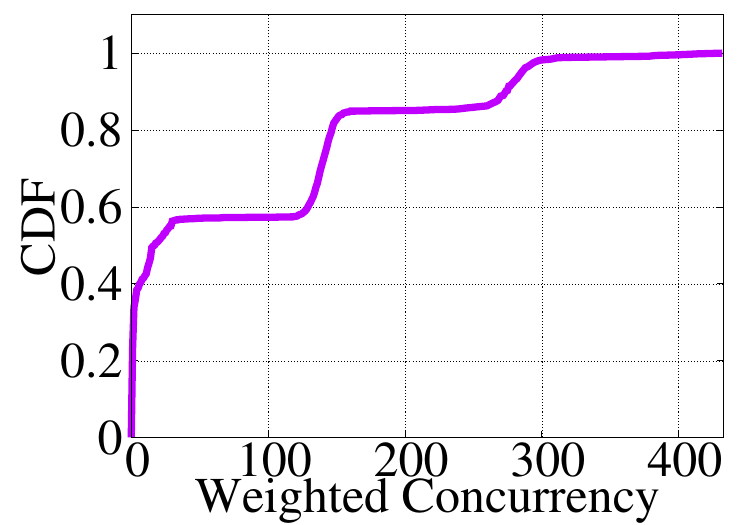}
    \footnotesize
  
    \textbf{(a) Weighted instance concurrencies of functions.}
  \end{minipage}
  \begin{minipage}[t]{0.49\linewidth}
    \centering 
    \includegraphics[scale=0.30]{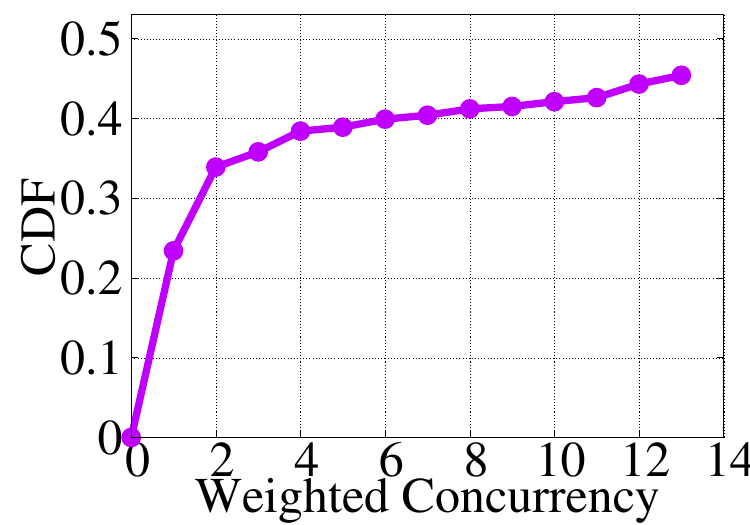}
    \footnotesize
  
    \textbf{(b) Weighted instance concurrencies (<13) of functions.}
  \end{minipage}
     \caption{\textbf{Analysis of \concurrency characteristic.}}
     %\caption{\textbf{Distribution of function concurrencies in public cloud.} Distribution of serverless instances (i.e., function * concurrency or weighted function) with concurrencies. Only 44\% instances are from functions whose concurrencies are less than 13.}
     % (a) Distribution of serverless functions with different concurrencies. (b) 
    \label{fig:trace-distribution}
  \end{figure}
  % \vspace{-0.8cm} 
% \myparagraph{Function concurrency distribution.}
\myparagraph{Serverless features that help tackle the challenge.}
We realize that the \emph{\concurrency} feature of serverless may help address the issue.
The feature can be illustrated with the following statistical analysis on real-world traces collected from \hwcloud's production environment.
Specifically, we mainly analyze the concurrency of function instances, i.e., the number of concurrently running instances of a function in a certain time window, rather than the invocation concurrencies in previous studies~\cite{10.1145/3503222.3507750,shahrad2020serverless,10.1145/3477132.3483580,10.1145/3567955.3567960}.
We weight the ratio of every concurrency value by itself, e.g., if there are 100 functions that have only 1 concurrency but 1 function whose concurrency is 100, the CDF point should be (1, 0.5) and (100, 1). 
% Some hot functions will have many instances which should have a higher weight in the distribution. % (we called it dynamic distribution).
The distribution is shown in Figure~\ref{fig:trace-distribution}-a.
Each point (e.g., ($x$,$y$)) in the figure indicates that ($y$$\times$100)\% instances belong to functions whose concurrencies are $\le x$.
%if a function has higher concurrency in a minute, there could be more instances of the function in the cloud.
For example, 56\% of instances are from functions whose concurrency is $>$12, which is shown more clearly in Figure~\ref{fig:trace-distribution}-b.
Instances of functions with only one concurrent instance are just 23\% of all instances.
In conclusion, such analysis shows serverless's \emph{\concurrency} feature --- \textbf{most serverless functions have many replicated instances}.

\myparagraph{Introducing capacity for decoupling.}
The feature means, for a deployed instance on a server, it is likely that more instances of the same function would come in the future.
Moreover, considering locality when deploying instances has also been proven beneficial by plentiful prior works~\cite{10.1145/3470496.3527390,10.1145/3552326.3567496}.
Therefore, to decouple prediction and decision making, we can predict in advance whether the next incoming instance of \emph{existing functions} can be deployed.
With this idea, for every server, \sys calculates \emph{capacity} for each existing function on that server.
%The capacity here is a bool value 
\sys predicts in advance the function's new instance's performance after deployment, and sets the capacity value ``true'' if the predicted performance meets its QoS.
The capacities for all functions on a server form the server's \emph{capacity table}.
After that, when the scheduler is making scheduling decisions, it can judge QoS violation by simply checking the table without model inference. 
It is the scheduling ``fast path''.
In addition, it is only necessary to make predictions on the critical path if the function to which the incoming instance belongs is not in the capacity table.
It is the scheduling ``slow path''.

\subsection{Asynchronous Update}
Although introducing capacity could predict the QoS violation of new instances in advance, the deployment of a new instance could impose resource interference to its neighbors, possibly causing their QoS violations.
Therefore, it is necessary to introduce a \emph{validation} process, which predicts and checks whether neighbors will violate their QoS after deploying the new instance.
However, such validation introduces costly inference for cold starts if done before deployment, or the risk of QoS violation if done after deployment.

To mitigate the defect, \sys applies an \emph{asynchronous update} approach.
To avoid the cost, asynchronous update refines the way to calculate the capacity, predicting all colocated functions' performances, and setting the capacity ``true'' if the predicted performance of each function meets its own QoS.
In this way, after deploying the new instance, the predicted performances of either the new instance itself or the neighbors would not violate the QoS.
This prevents the validation from affecting the scheduling latency.

Furthermore, when a new instance is scheduled to a server, it triggers updating the capacity table. 
Now that the validation can be delayed, and this update can be done asynchronously, outside the scheduling critical path. 
By asynchronous update, the capacity table is always up-to-date with minimum overhead, ensure that the capacity table always reflects real time interference on servers when scheduling.

%\subsection{Concurrency-aware Scheduling for Load Spikes}
\subsection{Concurrency-aware Scheduling}
\begin{figure}[t]
    \begin{minipage}[t]{0.48\linewidth}
      \centering 
      \includegraphics[scale=0.19]{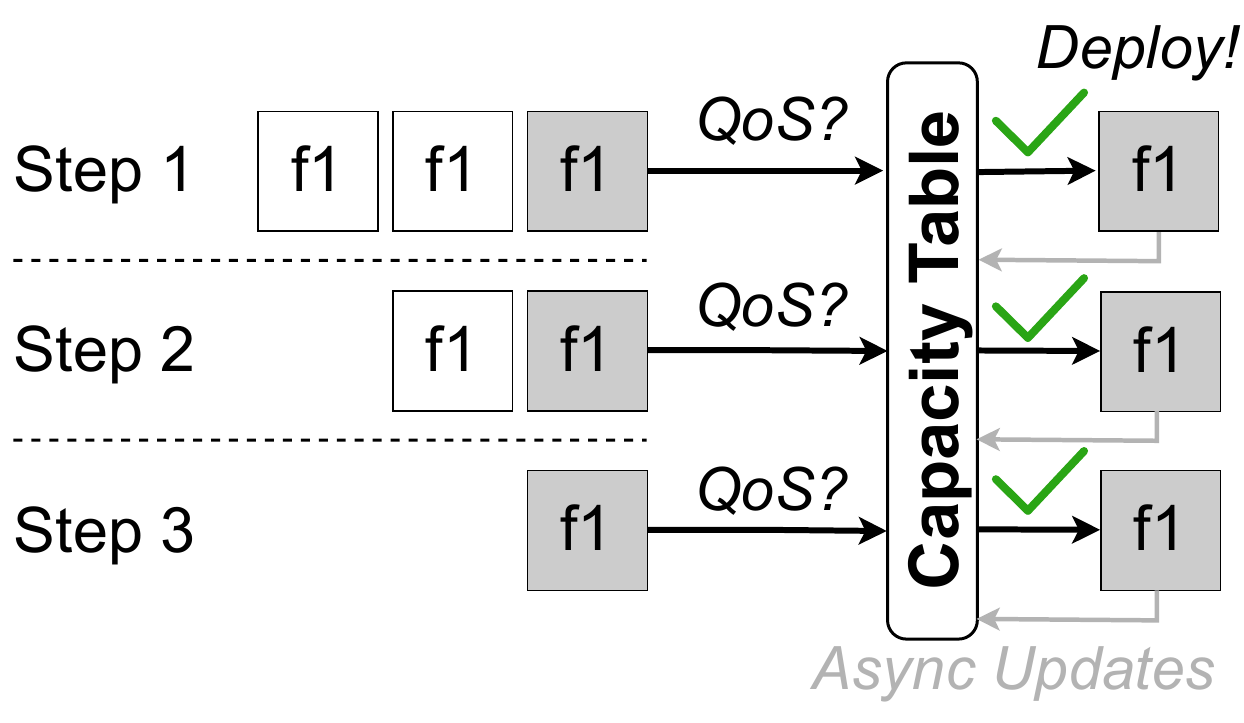}
      \footnotesize
    
      \textbf{(a) Without batching.}
    \end{minipage}
    \begin{minipage}[t]{0.48\linewidth}
      \centering 
      \includegraphics[scale=0.19]{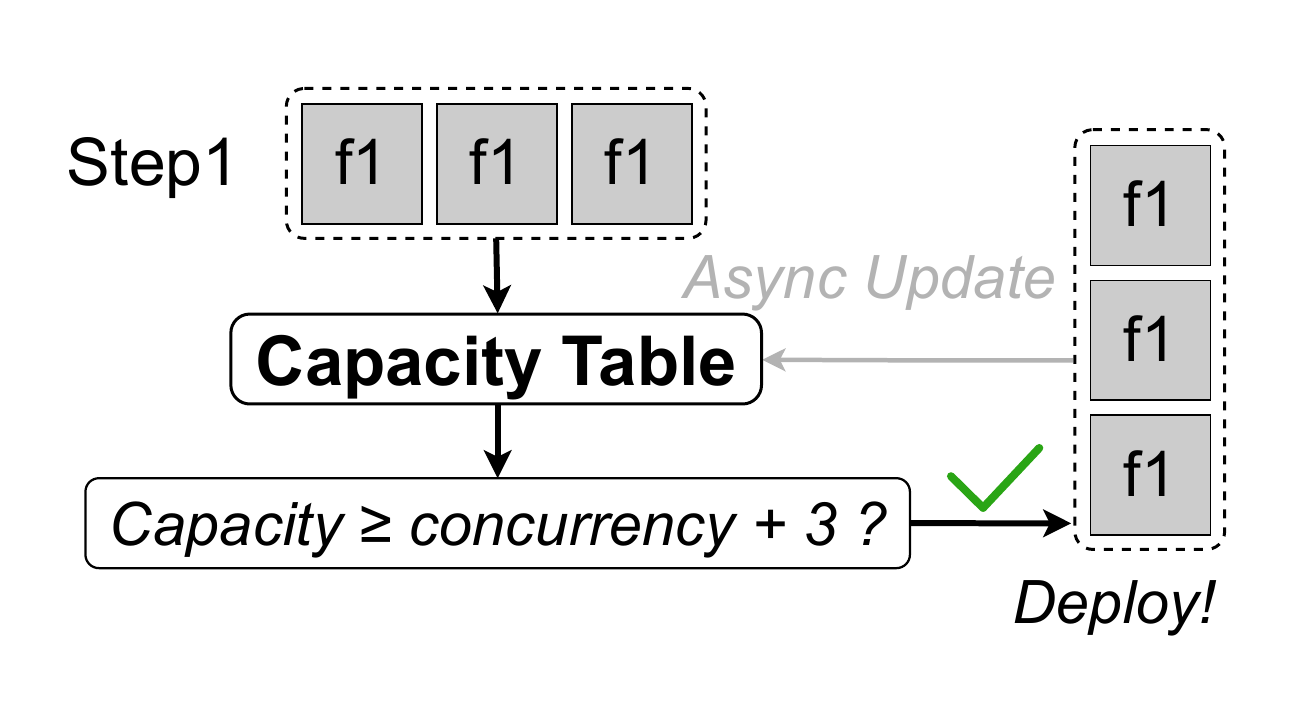}
      \footnotesize
    
      \textbf{(b) With batching.}
    \end{minipage}
    \caption{\textbf{Concurrency-aware scheduling enables batch scheduling for one function's multiple instances.}}
    \label{fig:capacity-batch}
\end{figure}

Decoupling prediction efficiently reduces model inference for individual schedulings. 
Moreover, one of the extreme invocation patterns in the cloud is the load spike, where the load rises rapidly so that multiple instances are created simultaneously. 
In such case, the asynchronous update of the previous instance could possibly block the scheduling of the next instance, as shown in Figure~\ref{fig:capacity-batch}-a.
To mitigate, \sys further adopts \emph{concurrency-aware scheduling}, which batches the scheduling of concurrent incoming instances at load spikes. 

\begin{figure}[t]
    \centering
    \includegraphics[width=0.42\textwidth]{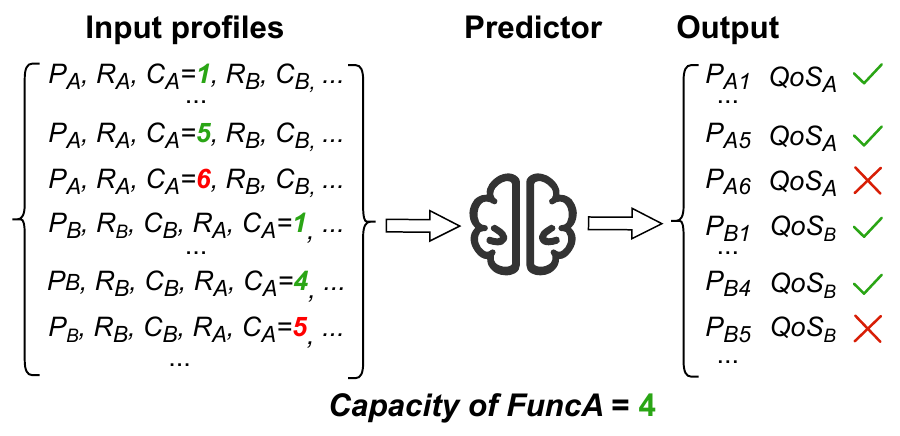}
	\caption{\textbf{Calculating capacity.} It finds the maximum concurrency value as the capacity which every colocated function's predicted performance can guarantee its QoS.}
    \label{fig:calculate-capacity}
  \end{figure}

\sys further refines the way of calculating capacity, considering not only whether the next instance can be deployed, but also how many next instances can be deployed, as shown in Figure~\ref{fig:calculate-capacity}.
The capacity value of each function is no longer a bool value, but a specific numerical value indicating how many instances of the function can be deployed with current neighbors. 
For example, for a server where there are 2 \fa ~and 3 \fb~instances, with \fa~and \fb's capacities are 4 and 6 respectively.
It means that 4 \fa~instances can be deployed with 3 \fb~instances, while 6 \fb~instances can be deployed with 2 \fa~instances.
Then, as shown in Figure~\ref{fig:capacity-batch}-b, when multiple instances of the same function arrive, if the function's capacity on the server is sufficient to accommodate those new instances, the scheduling and the asynchronous update can be done once for scheduling multiple instances.

\subsection{Put It All Together: Scheduling Example}
Now we describe the complete scheduling process and provide an example.
As shown in Figure~\ref{fig:design-overview}, when a function's new instance is created, \sys will first select a node for it using a node filter (detailed in \textsection\ref{s:impl}).
Then, based on the node's capacity table built in advance, the scheduler decides whether to schedule the instance via a fast path or a slow path. 

\begin{figure}[t]
    \centering
    \includegraphics[width=0.48\textwidth]{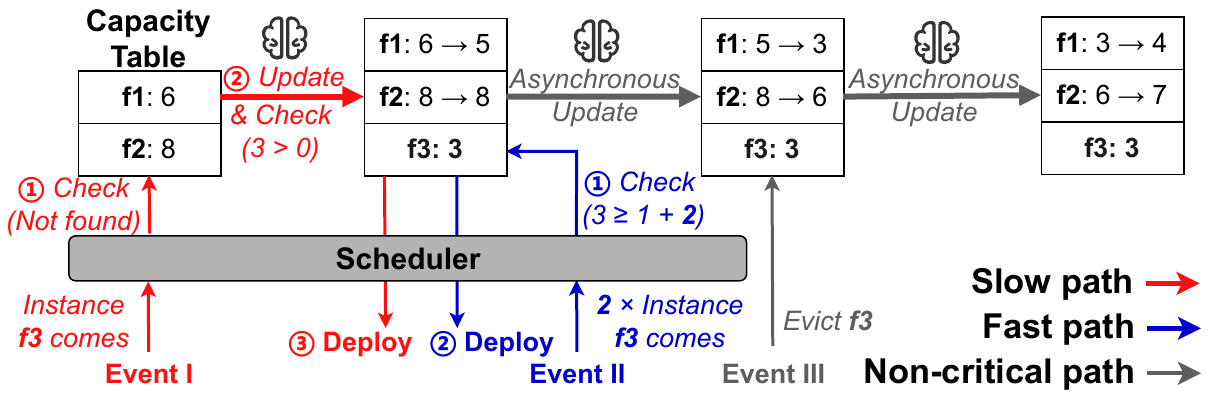}
	\caption{\textbf{Scheduling example on a node.}} % Dash circled instances are \textit{f1} and \textit{f2}'s scheduling results.}
    % Blue dash arrow shows the asynchronous update on \textit{Node 2} when \textit{f2} is scheduled to it.
    \label{fig:design-scheduler}
  \end{figure}

Figure~\ref{fig:design-scheduler} demonstrates an example.
First, an instance of \fc~incomes and the node filter selects the node in Figure~\ref{fig:design-scheduler} for it. 
The scheduler checks the capacity table and finds that there is no entry for \fc. 
Therefore, the scheduler calculates \fc's capacity using the prediction model (as shown in Figure~\ref{fig:calculate-capacity}). 
The result means that 3 \fc's instances can be deployed under current numbers of \fa ~and \fb ~instances, so the new instance of \fc~can be deployed on that server. 
It is the scheduling slow path involving a costly model inference.

Then, when load spike occurs, two instances of \fc~are created and come to the node. 
The scheduler checks the capacity table that the number of \fc's instances will not exceed its capacity after deployment, so that the scheduling decision is made quickly without model inference.
This is the scheduling fast path.
The deployment also triggers an asynchronous update of the capacity table, which can be done outside the scheduling critical path. 
The scheduling of two instances are batched, triggering only one update of the capacity table.

When load drops and an instance of \fc~is evicted, such event will also trigger an update to the capacity table. 
The increased capacities of \fa~and \fb~mean that the resources can be reutilized by the scheduler to deploy more instances later. 
Of course, this update would hardly affect the scheduling.

\section{Dual-staged Scaling}
\label{s:load-consolidation}
  \begin{figure}[t]
    \begin{minipage}[t]{0.48\linewidth}
      \centering 
      \includegraphics[scale=0.40]{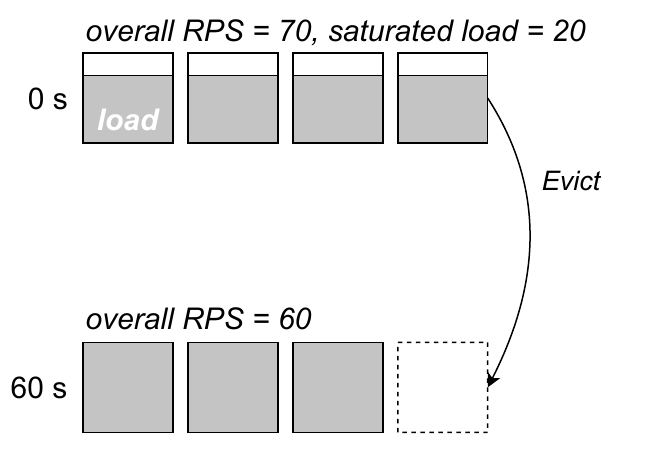}
      \footnotesize
    
      \textbf{(a) Existing eviction.}
    \end{minipage}
    \begin{minipage}[t]{0.48\linewidth}
      \centering 
      \includegraphics[scale=0.40]{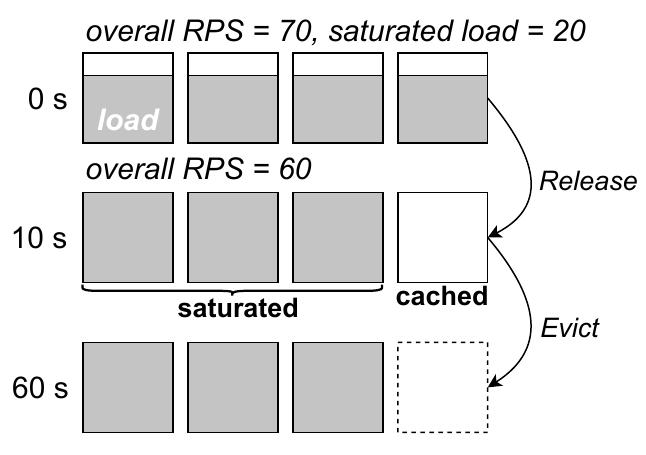}
      \footnotesize
    
      \textbf{(b) Dual-staged eviction.}
    \end{minipage}
    \caption{\textbf{Dual-staged scaling.} In the example, the \release and keep-alive duration are 10 and 60 seconds respectively.}
    \label{fig:design-load-consolidation}
\end{figure}

\iffalse old background in design
% It is well-known that serverless platforms usually apply a keep-alive policy~\cite{shahrad2020serverless, fuerst2021faascache}, which will delay the eviction of instances when users' load decreases to reduce cold starts in the near future. %(within the keep-alive duration).
% \sys's introduction of capacity ensures that even if instances are all \textit{saturated}, the QoS can be guaranteed when the function's current concurrency is lower than the capacity.
% However, as shown in the top two subgraphs of Figure~\ref{fig:design-load-consolidation}, when the load decreases, instances may no longer be saturated, freeing resource space for more instances of other functions. 
% The overlong keep-alive duration will waste this part of resources.
% To use the resources directly, an intuitive way is to weaken the keep-alive policy, evict instances immediately when the load drops, and update the concurrencies right away.
% Nevertheless, by doing so, the number of cold starts will significantly increase and harm the platform's performance, as shown in previous work~\cite{shahrad2020serverless,fuerst2021faascache,10.1145/3503222.3507750}.
\fi 

%\sys also proposes load consolidation to prevent the possible waste of resources caused by the fluctuations in user load.
%\highlight{
Motivated by the second insight, \sys designs dual-staged scaling to efficiently utilize resources under load fluctuation while minimizing cold start overheads.
%The design can be described by the following four parts.
%}

%\highlight{
\myparagraph{Dual-staged eviction.}
The key technique of the scaling approach is dual-staged eviction.
When load drops and so as the expected number of instances decreases, instead of directly evicting instances after the keep-alive duration as shown in Figure~\ref{fig:design-load-consolidation}-a, \sys introduces a ``\release'' duration (Figure~\ref{fig:design-load-consolidation}-b).
The ``release'' duration is (much) shorter than the keep-alive duration, i.e., the ``release'' operation is more sensitively triggered before eviction.
After the expected number of instances drops for the \release duration, the autoscaler first triggers the ``release'' operation, which changes the routing rules, sending requests to fewer instances, but does not actually evict any instances.
The instances that do not actually process requests are called \cached instances, while other instances that are still processing requests are called \saturated instances,
%(even though they may not actually be saturated),
as shown in Figure~\ref{fig:design-load-consolidation}-b.
%\highlight{
In the example, the load of four instances are consolidated to three instances. 
Therefore, when considering resource interference, the scheduler would consider only three \saturated instances instead of four, thus reducing resource wastage caused by overestimation of loads.
%}
After the ``release'' operation, the decrease in the number of \saturated instances will also trigger an asynchronous update and may increase the capacities of other functions on the server.
It means that the resources can be reutilized by the scheduler to deploy more instances.
%}

\myparagraph{Logical cold start.}
If the load rises again and the expected number of instances exceeds the current number of \saturated instances, if there are \cached instances at this point, \sys triggers a ``logical cold start''.
Specifically, \sys's scheduler chooses a \cached instance, then the router re-routes requests to the selected instance, letting it return to a \saturated instance.
This ``logical cold start'' is fast since the re-routing only takes $<1ms$, while a real instance initialization is far more costly.
% (\textsection~\ref{s:motiv}).

\myparagraph{On-demand migration.}
If the number of saturated instances of a function on a server drops and new instances of other functions are deployed to the server, the function's capacity could be updated and reduced.
Thus, after a while, the server could possibly be ``full'', i.e., the conversion of the function's \cached instances to \saturated instances on that server could result in the number of saturated instances surpassing its capacity.
At this moment, if the load rises again, \sys has to re-initialize instances on other feasible nodes, rather than merely trigger ``logical cold starts''.
To avoid incurring such additional ``real cold starts'', \sys learns in advance how many \cached instances are no longer converted to saturated instances by comparing the capacity with the sum of the numbers of both types of instances.
Then, it migrates these \cached instances to other feasible servers in advance.
Therefore, such additional cold start overheads can be hidden right away and will not affect the running of instances.

%\highlight{
\myparagraph{Real eviction.}
If the load does not rise again after the initial drop, finally after the keep-alive duration, the \cached instances are actually evicted.
After a while, if the load rises again and there are no \cached instances, it will trigger the autoscaler to initialize new instances, which involves ``real cold starts''.
It is consistent with the traditional autoscaling approach.

\section{Implementation}
\label{s:impl}

%\subsection{OpenFaaS Implementation}
%We present the prototype of \sys, based on the open-source serverless platform, OpenFaaS~\cite{openfaas}.
We implement \sys based on OpenFaaS~\cite{openfaas}, with 4.5k+ LoCs changes.
\sys is also developed and discussed at \hwcloud.

\iffalse old implementation
% When load fluctuates, \sys inquiries the controller and labels specific instances ``idle'' accordingly.
% Profiling worker processes run on separated servers, generating new data for the prediction model, and the prediction model does periodic online incremental updates with the new data to enhance prediction accuracy. 
% \sys's prototype runs the profiling workers as seperated processes on profiling nodes and training nodes.
% The workers deploy solo-run instances on profiling nodes or various colocation combinations on training nodes, collect profiling/performance metrics, and update the model used by the Kubernetes plugin.
% Our prototype on OpenFaaS has 4.5k+ LoCs changes.
\fi

\myparagraph{OpenFaaS architecture.}
OpenFaaS is a high-level controller based on Kubernetes (K8s), which schedules instances with the K8s scheduler and deploys instances with K8s resources (e.g., K8s \textit{Deployment} and \textit{Service}~\cite{k8s-service,k8s-deployment}).
It receives user requests from its gateway, and makes instance creation/eviction decisions with its autoscaling logic.
OpenFaaS relies on Prometheus~\cite{prometheus} to monitor the RPS of user requests and autoscale instances accordingly.
The autoscaler calculates the expected number of instances of a function by dividing the overall RPS by the predefined saturated load.
By default, it applies a 60-second keep-alive duration.
%according to the RPS.

\myparagraph{\sys's scheduling controller.}
%\sys prototype implements a plugin for the Kubernetes scheduler, allowing for interaction between the scheduler and the controller and enabling the scheduler to obtain placement decisions for incoming instances.
%It also extends Prometheus to implement load consolidation accordingly.
\sys's scheduling logic is implemented as a centralized controller, which manages the model, capacity tables,and other information required by the scheduling.
It works together with a plugin for Kubernetes scheduler in the prototype, allowing for interaction between the scheduler and the controller and enabling the scheduler to obtain placement decisions for incoming instances.
It also implements asynchronous update, where the scheduling results are returned first before performing the update.
Moreover, it is responsible for choosing instances for ``logical scaling'' for dual-staged scaling with its scheduling algorithm.

%\myparagraph{Implementing load consolidation and delayed eviction.}
\myparagraph{Implementing dual-staged scaling.}
To implement dual-staged scaling, \sys adds new rules for Prometheus to monitor RPS, and calculates the expected number of saturated instances according to the monitored values.
It applies a 45 (or 30, this sensitivity can be configured) seconds ``release'' duration, i.e., if the expected number of saturated instances decreases and last for 45 (or 30) seconds, it notifies the router for changing its routing rules.
To re-route, the scheduling controller chooses an instance and labels it as ``\cached''.
The ``\cached'' instances are excluded by the function's corresponding K8s \textit{Service}~\cite{k8s-service} so that new requests will not be routed to them.
When load rises and the expected number of saturated instances increases, it immediately triggers a logical cold start, i.e., the scheduler chooses a \cached instance and unlabels it.
%To implement delayed eviction, we still use the original autoscaler to decide when to evict an instance, with relatively low sensitivity.

\myparagraph{Profiling and model training.}
% \sys also implements an offline profiler to profile the functions and collect samples to construct the training set.
\sys adopts a \textit{solo-run} approach motivated by Gsight~\cite{10.1145/3458817.3476215} to collect the profile metrics of functions.
To profile a function, \sys addtionally deploys an instance to the profiling node exclusively, executes the function by the instance under the saturated load, and profiles the resource utilizations using tools such as \textit{perf} for a duration.
Moreover, to construct and maintain a training dataset, \sys further collects the performance metrics of various colocation combinations at runtime.
% \TODO{Incremental update here? clone requests here?}

\myparagraph{User configurations for functions.}
How users properly configure their functions~\cite{280890,288784} (e.g., configure resources and the saturated value) is out of the scope of the paper.
Even if a user can wisely configure their functions, the traditional approach of scheduling and autoscaling could still suffer from interference and load fluctuations (shown in Figure~\ref{fig:util-motiv}), leading to resource under-utilization.

\myparagraph{Node filter and management.}
The node filter described in Figure~\ref{fig:design-overview} prioritizes worker nodes in \sys.
For a new instance, it grants higher priority to nodes that have deployed instances of the same function, as scheduling on those nodes possibly follows a fast path.
While the current implementation focuses on this simple priority scheme, future work may explore extending it with additional scheduling policies, such as placing multiple instances of the same function on a single node to benefit from locality~\cite{10.1145/3470496.3527390,10.1145/3552326.3567496}.
In cases where no usable nodes are available for an instance, \sys will request the addition of a new server to the cluster.
Similarly, an empty server will be evicted to optimize costs.

\myparagraph{Hyperparameter configuration.}
The implementation involves configuring several specific parameters, including the profiling time and the definition of model convergence.
Setting these parameters requires careful consideration of trade-offs based on the actual scenario.
For example, collecting more samples when new functions arrive may result in a more accurate model but also incur higher overhead.
In our evaluation, all parameters remain consistent and aligned with our practices in \hwcloud.

\myparagraph{Performance predictability in the cloud.}
% Although single request performance can be affected by many factors such as user input, in practice, serverless functions still exhibit a certain level of predictability in terms of tail latency.
% This is because tail latency already includes multiple requests, and therefore has already taken into account various inputs and scenarios.
% There are also additional common practices that can benefit predictability, such as limiting the size of input images in the front-end for image processing functions.
% Our design is motivated by real-world observations that users are prone to over-allocate resources so that the resource utilization has great potential to be improved(ref and Figure).
% Moreover, 
%\highlight{
%Our design is based on the same assumption as prior works~\cite{8005341,10.1145/3274808.3274820,10.1145/3458817.3476215,10.1145/3620678.3624645}, i.e., 
QoS violations are mostly predictable, especially when using tail latency as the QoS metric, which considers the overall performance of multiple requests over a period of time.
It is consistent with the observations in our production environment and prior works~\cite{8005341,10.1145/3274808.3274820,10.1145/3458817.3476215,10.1145/3620678.3624645}.
In \hwcloud's production environment, \sys keeps monitoring the QoS violation and adopts two ways to handle the incidental unpredictability.
First, the random forest model is naturally feasible for incremental learning, and we continuously collect runtime performance metrics and retrain the model periodically with the up-to-date training set in case the behavior of functions changes.
Second, if the prediction error does not converge after several iterations, \sys disables overcommitment and uses traditional conservative QoS-unaware policy to schedule the instances of the unpredictable function on separate nodes.
\sys can perform well and improve resource utilization in real world.

\section{Evaluation}
\label{s:eval}

\iffalse
%In the evaluation, we answer the following questions:
%\begin{itemize}[leftmargin=*,topsep=0pt]
%
%	\item How is \sys's prediction model's accuracy? (\textsection\ref{subs:eval-predict})
%
%	\item Can \sys achieve great scheduling performance compared with prior ML-based schedulers? (\textsection\ref{subs:eval-perf})
%
%	% \item How can \sys improve resource utilizations while guaranteeing QoS on \ssim? (\textsection\ref{subs:eval-util})
%	
%  \item How can \sys improve resource utilizations while guaranteeing QoS on real serverless systems? (\textsection\ref{subs:eval-openfaas})
%
%	% \item Why \sys make the design choices in the paper? (\textsection\ref{subs:eval-discuss})
%\end{itemize}
\fi

\subsection{Methodology}
\myparagraph{Experimental setup.}
We use a cluster of 24 machines for evaluation.
Each machine is equipped with an Intel Xeon E5-2650 CPU (2.20GHz 48 logical cores in total) and 128GB memory, running Ubuntu LTS 18.04 with Linux 4.15 kernel.
%One machine is the master machine, and the OpenFaaS control plane components (gateway, scheduler, \sys's controller, etc.) are deployed on it.
One machine is dedicated to OpenFaaS control plane components (gateway, scheduler, \sys's controller, etc.). % are deployed on it.
The machine is also responsible for sending requests and collecting QoS and utilization results.
The other machines are worker machines responsible for executing or profiling function instances.
% We set the CPU frequency governor to ``powersave'' so that the CPU frequency can be stabilized at the lowest level.
% Machines are connected through 1000Mb ethernet cards.

\myparagraph{Baseline systems.}
% In the evaluation, we configured two versions of \sys prototype with 30 seconds and 45 seconds sensitivity of load consolidation respectively.
We compare \sys with three baseline systems: Kubernetes, Gsight~\cite{10.1145/3458817.3476215} and Owl~\cite{10.1145/3542929.3563470}.
Kubernetes scheduler is one of the mostly used serverless scheuduler systems in production~\cite{kube-scheduler,OpenWhisk}.
Gsight~\cite{10.1145/3458817.3476215} and Owl~\cite{10.1145/3542929.3563470} represent state-of-the-art predictor-based and historical information-based serverless scheduling systems.
Since they do not open source the implementation, we implement comparable prototypes of them.
Moreover, we evaluate three versions of \sys. 
Jiagu-30 and Jiagu-45 denote prototypes with 30 and 45 seconds ``\release duration'' of dual-staged scaling respectively, while \sys-NoDS disables dual-staged scaling. 

\iffalse
We compare \sys with three baseline systems.
Firstly, we compare \sys with traditional Kubernetes scheduler, one of the most wildly-used serverless scheduling algorithms and the default scheduling algorithm for several serverless systems~\cite{kube-scheduler,OpenWhisk}, which is conservative, prioritizing resource configurations.
Secondly and thirdly, we compare \sys with state-of-the-art predictor-based, and historical information-based serverless scheduling systems, Gsight~\cite{10.1145/3458817.3476215} and Owl~\cite{10.1145/3542929.3563470}.
Since they do not open source the implementation, we implement comparable prototypes of them.
Moreover, we evaluate three versions of \sys. 
Jiagu-30 and Jiagu-45 denote prototypes with 30 and 45 seconds sensitivity of load consolidation respectively, while \sys-NoLC disables load consolidation. 
\fi
% Second, we compare \sys with Knative's default scheduler based on Kubernetes.
% Kubernetes pod can specify multiple scheduling requirements like the node name, and Kubernetes scheduler has several extension points like filtering and scoring to help the pod meets its requirements.
% Each extension point has its functionality, for example, filtering methods select capable servers, and scoring methods give each server a priority for the pod.
% There are several default implementations for each extender points.
% For example, for scoring methods, kubernetes scheduler consider image locality, the balance of node resources, etc.

\myparagraph{Scheduling effect metrics.}
To evaluate the scheduling effect, we compare \sys with our baseline scheduling algorithms by two metrics.
First is the \textit{QoS violation rate}, which is the percentage of requests that violate QoS in all requests to all functions.
The QoS constraint is chosen to be 120\% of the tail latency when the instance is saturated and suffers no interference, consistent with previous work~\cite{10.1145/3542929.3563470,10.1145/3458817.3476215,10.1145/3567955.3567960,10.1145/3297858.3304005} and in-production practice.
In the evaluation, we aim to achieve a QoS violation rate of less than 10\% so we predict the p90 tail latency accordingly.
Second is the \textit{function density}, higher density means better resource utilization.
We normalize traditional Kubernetes scheduler's function density to one, meaning that instances are deployed with exactly the amount of configured resources.
One of the goals of \sys is to increase function density ($>$1) as much as possible while achieving an acceptable QoS violation rate ($<$10\%).

\begin{figure*}[htb]
  \setlength{\abovecaptionskip}{0.2pt}
  \setlength{\belowcaptionskip}{-10pt}
  \begin{minipage}[t]{0.32\linewidth}
        \centering
   \includegraphics[width=0.98\textwidth]{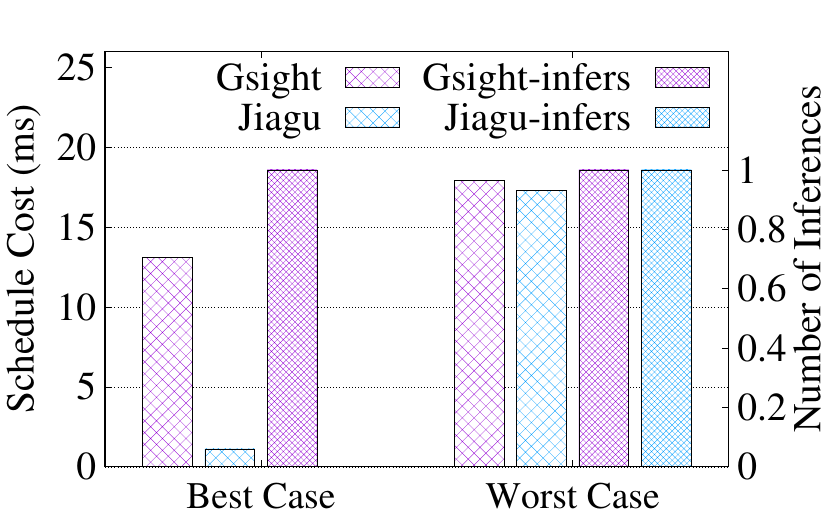}
    \footnotesize
    \textbf{(a) Scheduling cost and model inferences.}
\end{minipage}
\begin{minipage}[t]{0.32\linewidth}
    \centering
  \includegraphics[width=0.98\textwidth]{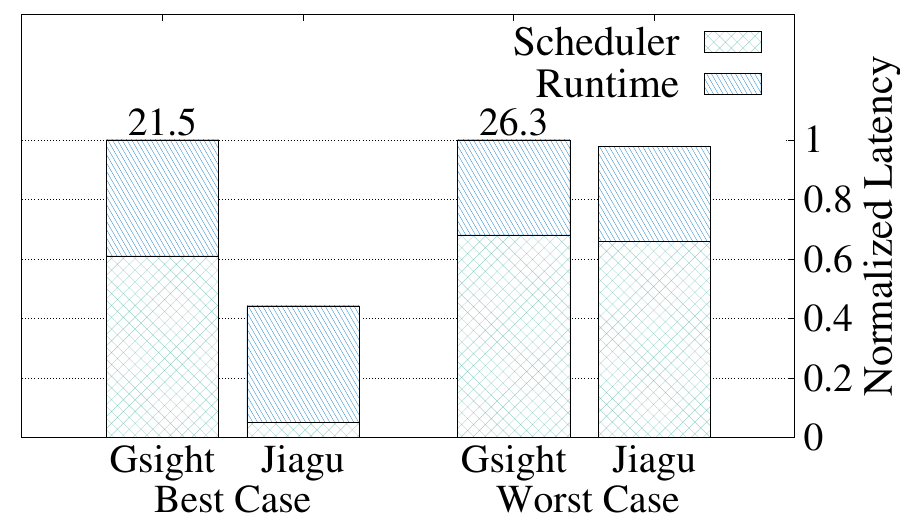}

    \footnotesize
    \textbf{(b) Cold start optimization with cfork.}
\end{minipage}
\begin{minipage}[t]{0.32\linewidth}
  \centering
\includegraphics[width=0.98\textwidth]{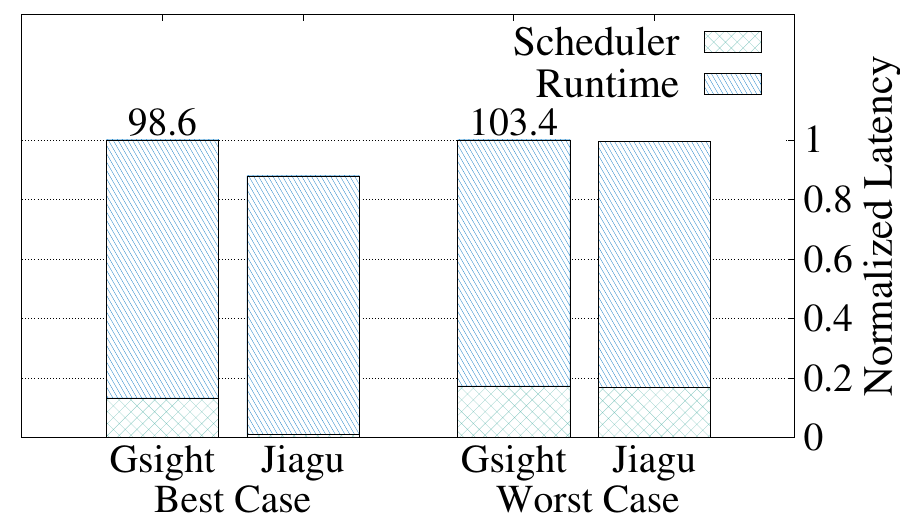}

  \footnotesize
  \textbf{(c) Cold start optimization with docker.}
\end{minipage}

\caption{\textbf{Performance analysis under extreme scenarios.} It includes scheduling cost, number of model inferences per schedule and cold start latency optimizations. We normalize results and label concrete numbers (ms) in the figure.}
\label{fig:eval-extreme-perf}
\end{figure*}

\begin{figure*}[htb]
  \setlength{\abovecaptionskip}{0.2pt}
  \setlength{\belowcaptionskip}{-10pt}
    \begin{minipage}[t]{0.32\linewidth}
        \centering
        \includegraphics[width=0.98\textwidth]{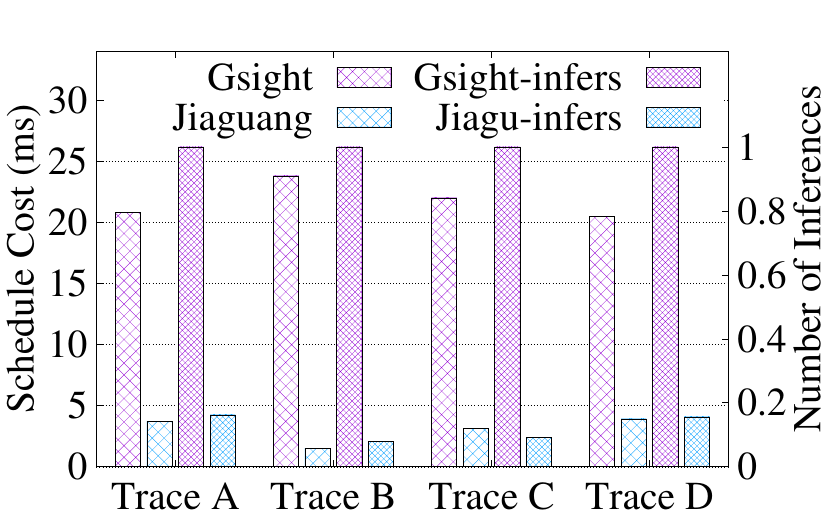}
        \footnotesize
        \textbf{(a) Scheduling cost and model inferences.}
    \end{minipage}
    \begin{minipage}[t]{0.32\linewidth}
        \centering
        \includegraphics[width=0.98\textwidth]{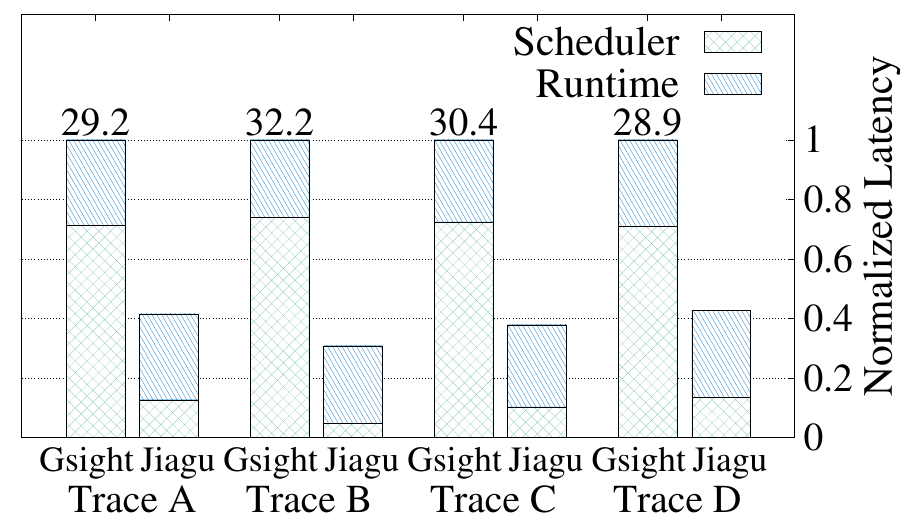}
        \footnotesize
        \textbf{(b) Cold start optimization with cfork.}
    \end{minipage}
    \begin{minipage}[t]{0.32\linewidth}
      \centering
    \includegraphics[width=0.98\textwidth]{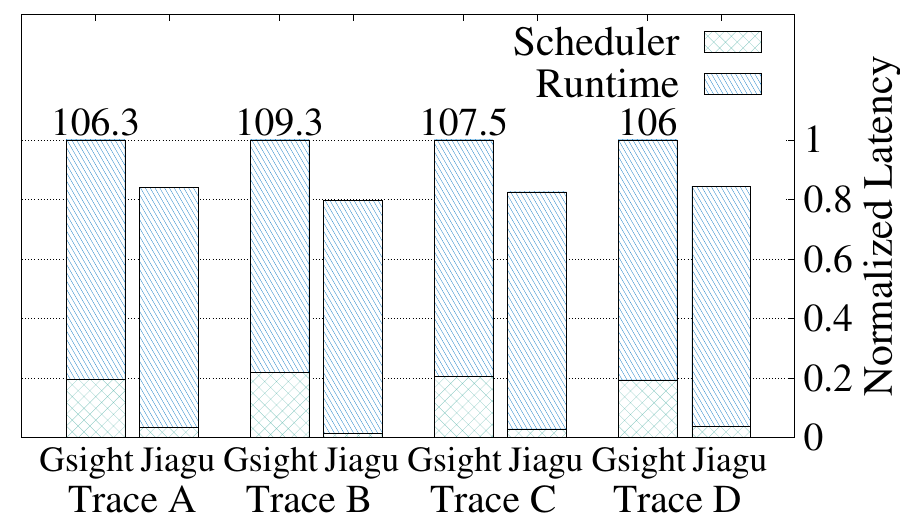}
    \footnotesize
      \textbf{(c) Cold start optimization with docker.}
  \end{minipage}
  
	\caption{\textbf{Performance analysis on real-world traces.} It includes scheduling cost, number of model inferences per schedule and cold start latency optimizations. We normalize results and label concrete numbers (ms) in the figure.}
    \label{fig:eval-sched-overhead}
\end{figure*}

\myparagraph{Workloads and traces.}
%The way of choosing the workloads and traces for our evaluation is consistent with prior work~\cite{10.1145/3542929.3563470,10.1145/3567955.3567960,10.1145/3458817.3476215,10.1145/3297858.3304005}.
The workloads and traces for our evaluation are consistent with prior work~\cite{10.1145/3542929.3563470,10.1145/3567955.3567960,10.1145/3458817.3476215,10.1145/3297858.3304005}.
We use six representative functions in ServerlessBench~\cite{socc20-serverlessbench} and FunctionBench~\cite{kim2019practical} for evaluation, including model inference (rnn), batch processing applications (image resize, linpack), log processing, chameleon and file processing (gzip).
All functions are configured with the same amount of resources.
To ensure the robustness of our system to perturbations in the benchmark inputs, we optionally add zero-mean Gaussian noise to the inputs.

We use various traces for evaluation.
First, we use real-world traces since they possess a similar level of complexity as in the production environment. 
For each set of real-world traces, we randomly select six invocation patterns from real-world \hwcloud traces and map each pattern to a function with similar execution time. 
We generate a total of four different sets of real-world traces to prevent the evaluation results from being influenced by randomness in the traces.
The four traces are also selected from real-world traces from various regions.
Second, we use specially constructed traces to evaluate the scheduling performance in extreme scenarios.

\subsection{Scheduling Performance Analysis}
\label{subs:eval-perf}

%This section evaluates \sys's performance using the model.

In this section, we evaluate the scheduling costs on OpenFaaS with different traces in two aspects.
First, we evaluate the average scheduling costs of the two algorithms.
Second, we simulate how \sys can optimize the cold start in conjunction with one of the state-of-the-art instance initialization optimizations, container fork (cfork)~\cite{10.1145/3503222.3507732}, which can create an instance in about 8.4ms. 
% Although \textit{cfork} overlooks the scheduling overhead in cold start latency, we show how \sys can further optimize cold starts in such scenarios.
We also analyze the cold start with Docker (about 85.5ms).
In each test, we compare \sys's default version (Jiagu-45) with Gsight, whose performance is normalized to 1.

\myparagraph{Performance under extreme scenarios.}
We first analyze the extreme performance of \sys using specific traces.
For the best case, we use a timer trace with only one function.
It assumes that the function is invoked and instances are scaled at a fixed frequency.
The results (Figure~\ref{fig:eval-extreme-perf}-a) show that Gsight suffers a scheduling overhead 11.9x larger than \sys. 
This is because almost all scheduling decisions of \sys go through the fast path in this case.
Considering cold start latency (Figure~\ref{fig:eval-extreme-perf}-b), Gsight's is 126.3\% longer than \sys when using cfork because of its costly scheduling.
% For the worst performance, we construct a rare trace where the function concurrencies recurrently change from 0 and 1, so that every scheduling process will incur a capacity calculation since every scheduled node is considered as an ``undeployed node''.
For the worst case, we construct a rare trace where the function concurrencies recurrently change from 0 and 1, so that every scheduling process will go through the slow path.
The results show that \sys's performance degraded to a similar level as Gsight.

When using Docker in both cases, the cost of instance initialization becomes the bottleneck.
We believe that as a growing number of efforts have limited initialization overhead to $<$10ms~\cite{DBLP:conf/usenix/WangCTWYLDC21, akkus2018sand, wang2019replayable,10.1145/3373376.3378512, oakes2018sock, jianightcore, 10.1145/3492321.3519581, 10.1145/3503222.3507732, cadden2020seuss, shillaker2020faasm}, the reduction in scheduling costs will be increasingly meaningful.

\myparagraph{Performance with real-world traces.}
We then analyze \sys's scheduling costs with four real-world traces. 
The results (Figure~\ref{fig:eval-sched-overhead}) show that for real-world traces, \sys achieves 81.0\%--93.7\% lower scheduling costs than Gsight.
This is because \sys's scheduling policy can drastically reduce the number of model inferences (83.8\%--92.1\%), so that the inference overhead is amortized over multiple cold starts.
For cold start latency, when applying \sys with cfork, the reduced scheduling costs lead to 57.4\%--69.3\% lower cold start latency than Gsight.

In conclusion, \sys's scheduling algorithm can significantly reduce model inference times and result in lower scheduling costs, leading to better cold start latency.

\begin{figure}[t]
  \setlength{\belowcaptionskip}{-10pt}
  \setlength{\abovecaptionskip}{-5pt}

  \begin{minipage}[t]{0.98\linewidth}
    \centering
    \includegraphics[width=0.94\textwidth]{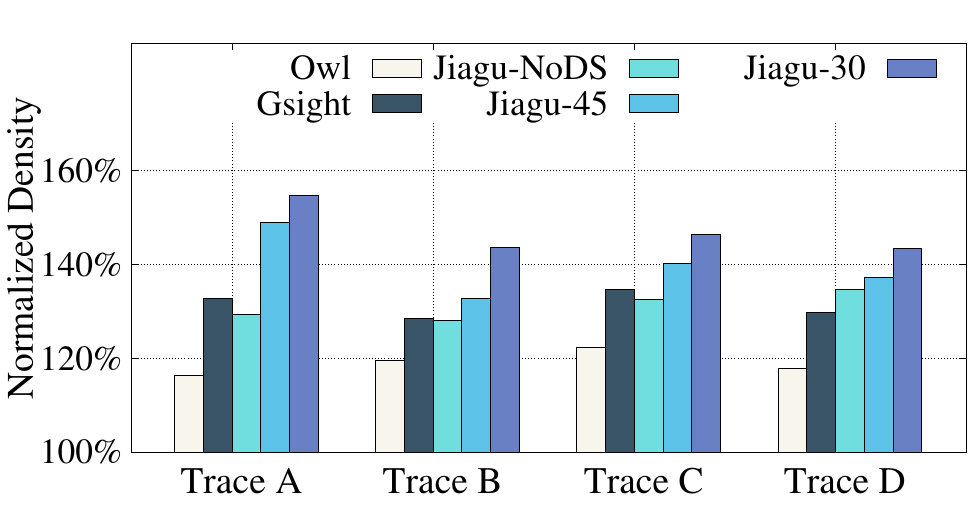}
    \footnotesize
    %\textbf{(a) Normalized function density.}
  \end{minipage}
  
  \caption{\textbf{Normalized function density.} Density on K8s is normalized to 100\%. \sys achieves the highest density with all of the real-world traces.}
  \label{fig:eval-of-eff}
\end{figure}

\begin{figure}[t]
   \setlength{\belowcaptionskip}{-10pt}
   \setlength{\abovecaptionskip}{0.5pt}
  \begin{minipage}[t]{0.52\linewidth}
    \centering
    \includegraphics[width=0.99\textwidth]{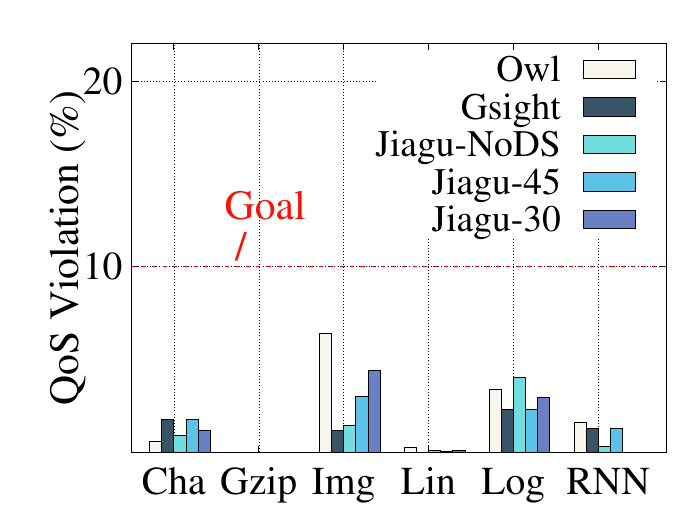}
    \footnotesize
    \textbf{(a) QoS violation of Trace A.}
  \end{minipage}
  \begin{minipage}[t]{0.47\linewidth}
    \centering
    \includegraphics[width=0.99\textwidth]{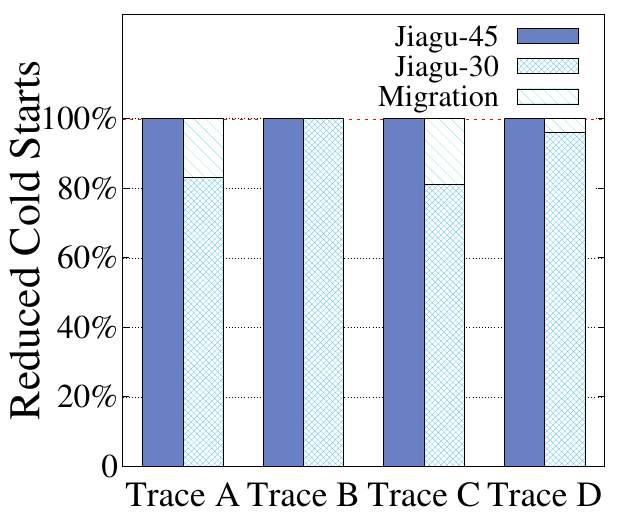}
    \footnotesize
    \textbf{(b) Reduced cold starts.}
  \end{minipage}
  \caption{\textbf{QoS violations and reduced cold starts.} Applications are chameleon (Cha), gzip, image resize (Img), linpack (Lin), log processing (Log) and RNN. }
  \label{fig:eval-delayed-eviction}
\end{figure}

\subsection{Scheduling Effect on OpenFaaS}
\label{subs:eval-openfaas}
% \myparagraph{Improving resource utilizations while guaranteeing QoS.}
This section describes how \sys improves resource utilization while guaranteeing QoS.
% We compare \sys with Gsight and a LoadBalance scheduling algorithm by two metrics.
% First is the \textit{QoS violation rate}, which is the percentage of requests that violate QoS in all requests to all functions.
% Second is the \textit{core efficiency}, which is the number of deployed instances' total requested cores to the number of actually allocated cores on the servers.
% Higher core efficiency means higher resource utilization.
% The goal of \sys is to achieve low QoS violation rate while increasing core efficiency as much as possible.
We analyze \sys and the baseline systems by the two metrics, function density and QoS violation rate, with the four real-world traces.
The density result is the average value during the evaluation weighted by duration.

According to Figure~\ref{fig:eval-of-eff}, all QoS-aware schedulers improve function density compared with Kubernetes (which is normalized to 1).
Without dual-staged scaling, Gsight and \sys (Jiagu-NoDS) achieve higher function density than Owl. 
It shows Owl's limitation of allowing only two colocated functions prevents optimal scheduling decisions.
With dual-staged scaling, \sys further optimizes resource utilization.
Higher sensitivity performs better in improving the instance density.
Specifically, Jiagu-30 achieves up to 54.8\% higher function density than Kubernetes, 22.0\% higher than Gsight and 38.3\% higher than Owl, while the QoS violation rate is comparable with Gsight and Jiagu-NoDS. 
It is because when user load drops, \sys can quickly utilize the resources of the unsaturated instances.
Moreover, on four traces, all schedulers achieve an acceptable QoS violation rate of less than 10\%, and the four results are similar.
Therefore, we only show the result on Trace A (Figure~\ref{fig:eval-delayed-eviction}-a).

\myparagraph{Reducing cold starts by migration of \cached instances.}
This section describes how many additional cold starts can be avoided by migrating \cached instances.
According to Figure~\ref{fig:eval-delayed-eviction}-b, for 45 seconds sensitivity, all re-routing operations are logical cold starts, needless to migrate instances.
With 30-second sensitivity, only a small proportion of re-routing, i.e., $<20\%$ for four traces and $0$ for Trace B, requires ``real'' rather than ``logical'' cold starts, which can be avoided by migrating \cached instances in advance.
It illustrates that with dual-staged scaling, \sys has the additional choice of greatly improving resource utilization at minimum cost.

In conclusion, with the prediction-based scheduler and dual-staged scaling, \sys can achieve high resource utilization while limiting QoS violations with less scheduling overhead than state-of-the-art model-based scheduling policy.

\begin{figure}[t]
  \begin{minipage}[t]{0.48\linewidth}
    \centering
    \includegraphics[width=0.98\textwidth]{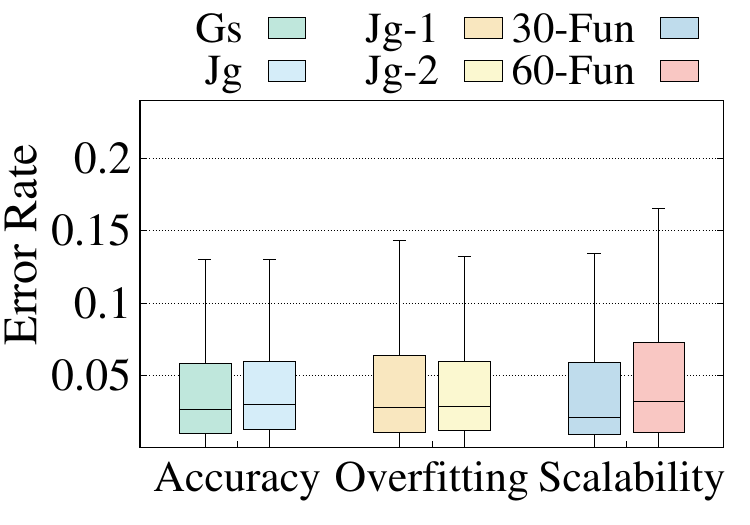}
    \footnotesize
    % \textbf{(a) Prediction error.}
  \end{minipage}
  \begin{minipage}[t]{0.50\linewidth}
    \centering
    \includegraphics[width=0.98\textwidth]{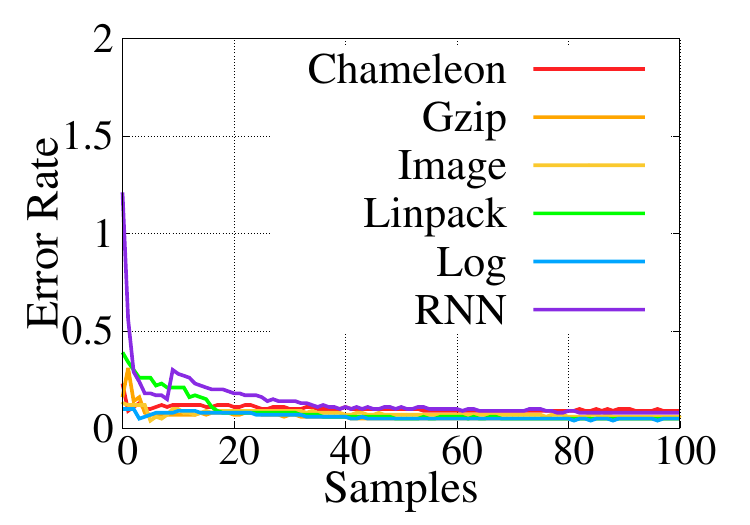}
    \footnotesize
    % \textbf{(a) Prediction error.}
  \end{minipage}
  \caption{\textbf{Model accuracy}.
(a) is the error rate of the prediction models.
(b) shows that the prediction error drops with new samples.
}
\label{fig:eval-model-accuracy}
\end{figure}

\begin{figure}[t]
  \setlength{\abovecaptionskip}{0.2pt}
  \setlength{\belowcaptionskip}{-10pt}
    \begin{minipage}[t]{0.98\linewidth}
    \centering
    \includegraphics[width=0.90\textwidth]{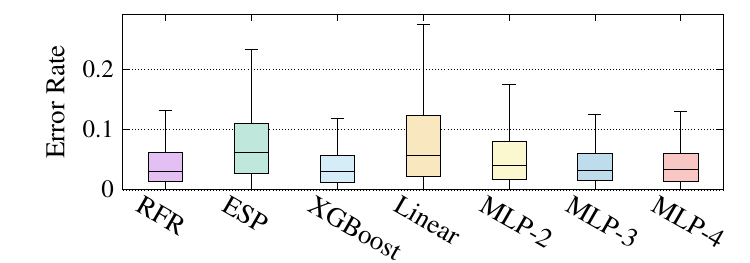}
    \footnotesize
    %\textbf{(a) Normalized function density.}
  \end{minipage}
  
  \caption{\textbf{Prediction errors of various models.}}
  \label{fig:eval-of-models}
\end{figure}

\begin{figure}[t]
  \setlength{\abovecaptionskip}{0.2pt}
  \setlength{\belowcaptionskip}{-10pt}
  \begin{minipage}[t]{0.49\linewidth}
    \centering
    \includegraphics[width=0.98\textwidth]{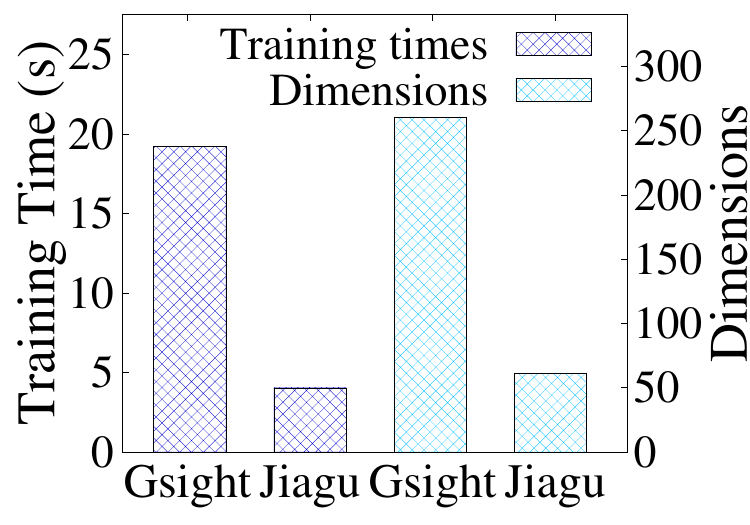}
    \footnotesize
    % \textbf{(c) Training overhead and dimensions.}
  \end{minipage}
  \begin{minipage}[t]{0.49\linewidth}
    \centering
    \includegraphics[width=0.98\textwidth]{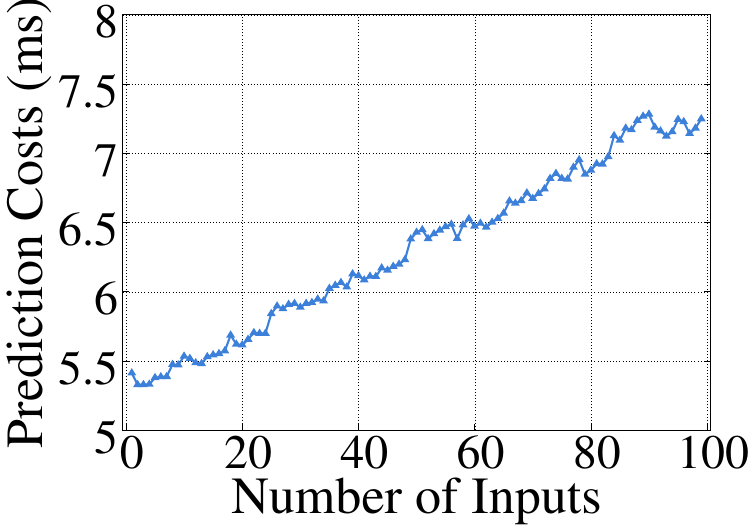}
    \footnotesize
    % \textbf{(c) Training overhead and dimensions.}
  \end{minipage}
  \caption{\textbf{Model performance.} (a) is the training time and the total number of dimensions of the two models.
  (b) \sys's model inference overhead with the number of inputs increases.}
  \label{fig:eval-model-perf}
\end{figure}

\subsection{Prediction Analysis}
\label{subs:eval-predict}
\myparagraph{Prediction accuracy.}
%\subsubsection{Question 1}:
% We first evaluate the prediction accuracy of our model and the training costs, using both \ssim and OpenFaaS.
We evaluate the prediction accuracy of our model, as shown in Figure~\ref{fig:eval-model-accuracy}.
We define the prediction error rate of a model as $\frac{|\hat{P} - P|}{P}$, where $\hat{P}$ and $P$ are predicted performances and real performances, respectively.
First, we analyze the accuracy of the model used by previous evaluations.
The model could accurately predict the performance of the six functions, with similar prediction error to Gsight (Jg and Gs in the Figure).
Second, we evaluate the overfitting of the model by splitting the test set into two equal-sized sets (Jg-1 and Jg-2).
The model achieves similar prediction error on the two different sets, showing that it does not overfit to a specific test set.
Third, we evaluate the model's scalability, i.e., how the model can predict the performance of increasing number of functions (30 and 60 functions).
The result shows that more functions do not affect the prediction precision.
Fourth, we evaluate how the model is resilient to the changes in function's behavior.
Each time, we train the model with five functions, and add the remaining function to the system, collect performance metrics as new samples and retrain the model once a sample is added.
% Each time we train the model with five functions, and treat the other function as a newly added function to the system, colocate it with various colocation combinations, and collect its performances as the test set. 
% Then, we colocate the function with new colocation combinations, collect the performances as new samples, retrain the model once a new sample is added, and evaluate the prediction errors of every retrained model on the test set.
Figure~\ref{fig:eval-model-accuracy}-b shows that the performance prediction error of the new function drops rapidly under increasing number of samples, and converges after about 5--30 samples. 
It means that it takes limited time to retrain an accurate model after collecting a couple of samples.
\iffalse
% Firstly, we profile and evaluate the model on OpenFaaS.
% To train our model, we select a small subset (3,438 samples, 80\% of them are training set and the rest are test set) of the entire space of the combinations of six functions selected from the benchmarks. 
% Each function has various manually constructed concurrencies.
% According to Figure~\ref{fig:eval-model-accuracy}(a), the prediction error of \sys's model is similar to that of Gsight. 

% Secondly, we evaluate the overfitting of the model.
% We randomly split the test set into two equal-sized sets (Jiagu-1 and Jiagu-2 in the figure) and evaluate the prediction error on both of them.
% According to Figure~\ref{fig:eval-model-accuracy}(a), the model achieves similar prediction error on the two different test sets.
% It shows that \sys's prediction model does not overfit to a specific test set.

% Thirdly, we evaluate the accuracy with a larger number of heterogenenous functions, 30 and 60 respectively.
\fi

Finally, we evaluate the model choices, comparing \sys's random forest regression (RFR) with ESP~\cite{8005341}, XGBoost, linear regression, and three multi-layer perception (MLP) models with 2,3,4 layers respectively.
The result (Figure~\ref{fig:eval-of-models}) confirms our choice of RFR model considering its high accuracy, low training overhead and capability of incremental learning, etc.
Possible other models or algorithms (e.g., ridge regression) can be explored to improve \sys in future work.

\myparagraph{Model performance.}
We evaluate \sys's model training and inference costs.
Figure~\ref{fig:eval-model-perf}-a shows that \sys's model has evidently better training performance and lower number of input dimensions.
Moreover, Figure~\ref{fig:eval-model-perf}-b demonstrates that the model inference costs increase only 2 milliseconds even when the number of inputs increases to 100.
Therefore, although \sys batches multiple inputs to the prediction model when calculating the capacity, it will not significantly increase the inference overhead.

In conclusion, \sys's model can predict the performance accurately with a much lower number of input dimensions and training overhead.

\section{Related Work}
\label{s:relwk}

\myparagraph{Resource schedulers.}
%\MZY{This paragraph should not be placed side by side with the above two paragraphs.}\DD{Move this part to related work now.}
% Another widely studied (and orthogonal) scheduler is the online resource scheduler~\cite{10.1145/3445814.3446700,10.1109/MICRO.2014.53,10.1145/2508148.2485974,10.1145/3330345.3330370,10.1145/2465351.2465388,10.1145/3297858.3304004,10.1145/2749469.2749475,10.1145/3297858.3304005,9065583,10.5555/3488766.3488812,10.1145/3445814.3446693,10.5555/3488766.3488782,10.1145/3472883.3486972,10.1145/2663165.2663330,10.1145/3567955.3567960,10.1145/3542929.3563469}, which will monitor metrics and predict the QoS during runtime, and tune the allocated resources accordingly to satisfy QoS requirements.
Another widely studied (and orthogonal) scheduler is the online resource scheduler~\cite{288542,10.1145/3342195.3387524,10.1145/3445814.3446700,9380428,10.1109/MICRO.2014.53,10.1145/2508148.2485974,10.1145/3330345.3330370,8457882,10.1145/2465351.2465388,10.1145/3297858.3304004,10.1145/2749469.2749475,10.1145/3297858.3304005,9065583,10.5555/3488766.3488812,10.1145/3445814.3446693,10.5555/3488766.3488782,10.1145/3472883.3486972,10.1145/2663165.2663330,9155363,10.1145/3567955.3567960,10.1145/3542929.3563469,10.1145/2806777.2806779,10.1145/2541940.2541941,8711058,9284283,9773201,cuttlesys}, which tunes the allocated resources according to functions' online performance to guarantee QoS and improve resource efficiency.
Many of them are designed for long running services~\cite{10.1145/3342195.3387524,10.1145/3445814.3446700,10.1109/MICRO.2014.53,10.1145/2508148.2485974,10.1145/3330345.3330370,10.1145/2465351.2465388,10.1145/3297858.3304004,10.1145/2749469.2749475,10.1145/3297858.3304005,9065583,10.5555/3488766.3488812,10.1145/3445814.3446693,10.5555/3488766.3488782,10.1145/3472883.3486972,10.1145/2663165.2663330,10.1145/3542929.3563469}, which are challenging to apply directly to transient serverless instances. 
In addition, some serverless resource schedulers~\cite{9155363, 10.1145/3567955.3567960,8457882} use algorithms like Bayesian Optimization to search the resource configuration space and make optimal decisions based on runtime performance. 
They are resilient to runtime performance variations but still fail to guarantee QoS before convergence.
% However, as interference in the cloud continuously changes, they have to continuously modify resource configurations, making it difficult to maintain a stable optimal status.

%Unlike instance schedulers, 
% Since applications' performance fluctuates over time, resource schedulers need to predict the QoS violation periodically and adjust resource allocation accordingly, which will incur significantly higher runtime costs. % compared with the above two instance schedulers. 
% \MZY{Why not profile offline?} 
% \LQY{Is it ok that just describe the cons of Resource scheduler in this section?}

%\myparagraph{Routers.}
\myparagraph{Request schedulers.}
Rather than scheduling instances, many prior works tackle the problem of assigning instances for requests.
Atoll~\cite{10.1145/3472883.3486981} overcomes many latency challenges in the serverless platform via a ground-up redesign of system control and data planes, utilizing deadline-aware scheduling.
Hermod~\cite{10.1145/3542929.3563468} analyzes and tunes different design choices on OpenWhisk and improves function performance. 
Fifer~\cite{10.1145/3423211.3425683} queues requests to warm containers to reduce the number of instances and prewarms instances with LSTM resource interference to optimize the cold starts. 
Sequoia~\cite{10.1145/3419111.3421306} is deployed as a proxy to control user requests and enabling QoS.
Neither of them considers resource interference.
%\highlight{
Golgi~\cite{10.1145/3620678.3624645} applies a model for the router, predicting requests' performance and sending them to overcommitted instances if the performance does not violate QoS, or otherwise to non-overcommitted instances.
It can avoid QoS violation for a fixed number of instances, but does not guide for how many overcommitted instances need to be deployed, while excessive number of overcommitted instances could cause resource under-utilization.
%}

\iffalse
\myparagraph{Other instance schedulers.}
Some instance schedulers effectively improve resource utilization for the monolithic service, but fail to adapt to the serverless scenario.
Bubble-up~\cite{10.1145/2155620.2155650} only considers the resource interference in the memory system.
Paragon~\cite{10.1145/2451116.2451125} profiles an application and scores its sensitivity and pressure to resources.
They can not accurately predict or even do not actually predict the QoS violation.
Moreover, Whare-map~\cite{10.1145/2485922.2485975} utilizes Google Wide Profile~\cite{36575} to exhaustively profile various colocation combinations and score each server accordingly.
%In serverless scenario, where the number of functions and the number of function instances can be large and constantly changing, it is impossible to profile such an abundant number of application combinations.
Pythia~\cite{10.1145/3274808.3274820} requires maintaining a model for each application.
Its profiling or training overhead can be almost impractical considering the number of functions can be infinite.
\fi

\myparagraph{Other optimizations.}
There are many related serverless optimizations~\cite{280890, 10.1145/3503222.3507750, 10.1145/3503222.3507717, 222563, 10.1145/3472883.3486972, 10.1145/3419111.3421306, 10.14778/3407790.3407836, 10.1145/3477132.3483541}.
For example, Orion~\cite{280890} proposes a system optimized for serverless DAG by proposing a new performance model and a method for colocating parallel invocations. %to consider runtime variabilities and dependencies among functions in a DAG 
FaasFlow~\cite{10.1145/3503222.3507717} proposes a worker-side workflow schedule pattern that enables better scalability and performance.
All of these systems can benefit from \sys with a better and more effective scheduler to also improve the resource utilization for cloud vendors without introducing high scheduling costs.

%\paragraph{Open-sourced serverless systems.}
% Existing open-sourced serverless systems still use simple schedulers.
% For example, OpenWhisk~\cite{OpenWhisk} uses load balancing to co-locate instances of the same function to a randomly-selected node.
% It does not consider load as well as QoS.
% OpenFaaS~\cite{openfaas} and vHive~\cite{10.1145/3445814.3446714} are designed based on Kubernetes, and rely on an auto-scalers to monitor the CPU and memory utilization and launch or terminate instances when the utilization changes.
% The reactive approach can not handle serverless workloads' dynamicity and short life-cycle.

%Resource schedulers can also satisfy the goal of a good serverless scheduler but will incur higher runtime costs and are hard to deploy to commercial serverless systems with dynamic machines (i.e., node auto-scalability) and fixed rules to manage resources.

\section{Conclusion}
\label{s:conclusion}

This paper presents our experience in optimizing resource utilization in \hwcloud for serverless computing with \sys.
%\sys's design is inspired by serverless's \concurrency nature.
\sys decouples prediction and decision making to achieve efficient and fast scheduling, and decouples resource releasing and instance eviction to achieve fast reaction to load fluctuation without incurring overmuch cold start overhead.
Our results show \sys brings significant benefits for resource efficiency using real-world traces.

%real-world applications.
%We believe \sys will motivate many future works for serverless schedulers.

% \input{async-start}
% \input{declarative-tax}
% \input{sched-cost}
% \input{sched-policy}
% \input{sidecar}
% \input{concl}

% \input{ack}

\balance
%%%%%%% -- PAPER CONTENT ENDS -- %%%%%%%%

%-------------------------------------------------------------------------------
% \bibliographystyle{plain}

\bibliographystyle{ACM-Reference-Format}
\bibliography{ref}

\end{document}
\endinput
%%
%% End of file `sample-sigconf.tex'.